\pdfoutput=1
\documentclass[aps,prd,superscriptaddress,showkeys,onecolumn,nofootinbib,10pt]{revtex4-2}
\usepackage{preamble}
\begin{document}

\title{Introduction to the number of $e$-folds in slow-roll inflation}

\author{Alessandro Di Marco\orcidA{}}
\email{alessandro.dimarco1@inaf.it}
\affiliation{Istituto Nazionale di Astrofisica,\\
Istituto di Astrofisica e Planetologia Spaziali (INAF-IAPS),\\
Via Fosso del Cavaliere, 100, 00133 Rome, Italy}

\author{Emanuele Orazi\orcidB{}}
\email{orazi.emanuele@gmail.com}
\affiliation{Escola de Ci$\hat{e}$ncia e Tecnologia and International Institute of Physics,
Federal University of Rio Grande do Norte, Campus Universit´ario-Lagoa Nova, Natal-RN 59078-970, Brazil}

\author{Gianfranco Pradisi\orcidC{}}
\email{gianfranco.pradisi@roma2.infn.it}
\affiliation{University of Rome ``Tor Vergata'' \\
and INFN, Sezione di Roma ``Tor Vergata'', via della Ricerca Scientifica 1, 00133 Roma, Italy}

\date{\today}

\begin{abstract}
In this review, a pedagogical introduction to the concepts of slow-roll inflationary Universe 
and number of $e$-folds is provided.
In particular, the differences between the basic notion of number of $e$-folds ($N_e$),
total number of $e$-folds ($N_T$) and number of $e$-folds before the end of inflation ($N$), are outlined. 
The proper application of the number of $e$-folds before the end of inflation 
is discussed both as a time-like variable for the scalar field evolution and as a key parameter for computing inflationary predictions.

\end{abstract}

\keywords{Cosmology; Early Universe cosmology; Inflation; Number of $e$-folds}
\maketitle
\tableofcontents

\section{Introduction}
\label{introduction}

The paradigm of cosmological inflation 
\cite{Starobinsky:1980te, Guth:1980zm,Linde:1981mu,Albrecht:1982wi,Hawking:1981fz,Linde:1983gd}
(for specific reviews, 
see~\cite{Linde:1990flp,Linde:2007fr,Olive:1989nu,Baumann:2009ds},
and for a pedagogical wide-ranging treatment, see~\cite{Kolb:1990vq,Mukhanov:2005sc,Weinberg:2008zzc,Gorbunov:2011zzc,Baumann:2018muz})
offers a comprehensive solution for understanding the origins of the specific initial 
conditions that underpin the standard Hot Big Bang (HBB) theory. 
These conditions encompass the flatness of three-dimensional constant-time hypersurfaces, 
the homogeneity and isotropy of the cosmic microwave background (CMB), 
the shortage of hypothetical heavy relics, and the significant observable entropy. 
Moreover, inflation generates adiabatic, Gaussian, and nearly scale-invariant scalar metric perturbations that are responsible both for matter inhomogeneities leading to the formation of the
observable large-scale structures (galaxies, clusters of galaxies, superclusters, etc.)
and for the primary (relative) temperature anisotropies ($\sim$10$^{-5}$) of the \mbox{CMB 
\cite{Mukhanov:1981xt,Hawking:1982cz,Starobinsky:1982ee,Guth:1982ec,Linde:1982uu,Bardeen:1983qw,Lyth:1984gv,Mukhanov:1985rz,Sasaki:1986hm,Stewart:1993bc}.}
Nevertheless, inflation gives rise to tensor perturbations, or primordial gravitational waves (GWs), 
which could be detectable if the inflationary energy scale is sufficiently high
\cite{Grishchuk:1974ny,Starobinsky:1979ty,Rubakov:1982df,Fabbri:1983us,Abbott:1984fp,Allen:1987bk,Lucchin:1992qi,Turner:1993vb,Crittenden:1993wm,Knox:1994qj,Turner:1996ck}.
For complete reviews on scalar and tensor perturbations, \mbox{see~\cite{Mukhanov:1990me,Riotto:2002yw,Guzzetti:2016mkm}}.
The inflationary phase can be realized via several mechanisms, 
and the simplest one is the so-called (single-field) slow-roll inflation~\cite{Steinhardt:1984jj,Liddle:1992wi,Liddle:1994dx} 
that exists in a large number of varieties~\cite{Martin:2013tda}.
The minimal version of the slow-roll scenario involves a
single, homogeneous, neutral, minimally coupled, and canonically normalized 
(pseudo)scalar field $\phi$, 
known as the \textit{inflaton}, which typically dominates 
the stress--energy tensor $T_{\mu\nu}$ 
of a {reliable} local universe's \textit{patch} at (or below) the Planck scale 
(see~\cite{Coughlan:1985mw,Linde:1985ub,Albrecht:1984qt,Albrecht:1985yf,Albrecht:1986pi,Kurki-Suonio:1987mrt,Goldwirth:1989pr,Goldwirth:1991rj,Kurki-Suonio:1993lzy,Iguchi:1996rh,Vachaspati:1998dy,Easther:2014zga,East:2015ggf} for details on such a slow-roll onset).
On general ground, the inflaton's scalar potential $V(\phi)$, referred to as the \textit{inflationary potential},
exhibits an almost flat region and a global vacuum.
Initially, the inflaton field is misaligned from the potential minima and slowly crosses the flat region. 
Consequently, the potential contribution dominates over the field kinetic term,
resembling the presence of a false vacuum or a transient cosmological constant and 
inducing an almost de Sitter expansion of \mbox{the universe.}

After traversing a total distance $\Delta\phi$ 
\cite{Lyth:1996im,Efstathiou:2005tq,Easther:2006qu,Garcia-Bellido:2014eva,Garcia-Bellido:2014wfa,DiMarco:2017ihz}
and reaching a value $\phi_{\text{end}}$, the slow-roll of the field breaks down and the accelerated phase ends. 
The inflationary expansion yields a universe that is approximately flat, smooth, empty, and cold, as all pre-existing energy and numerical densities are diluted away. 
Nevertheless, as $\phi$ approaches $\phi_{\text{end}}$, 
the kinetic term becomes important and
the inflaton rapidly drops to the true vacuum, around which it begins to oscillate.
The inflaton field must also be coupled to the degrees of freedom of the Standard Model (SM) or 
of some Beyond the Standard Model (BSM) extensions,
enabling the transfer of energy density stored in the scalar oscillation to the SM (or BSM) component fields. 
This phase is called \textit{reheating}. It produces the relativistic plasma of light SM (or BSM) particles
and the corresponding large entropy, and triggers the radiation epoch of the Hot Big Bang (HBB) cosmology
(see
\cite{Albrecht:1982mp,Dolgov:1982th,Abbott:1982hn,Turner:1983he,Shtanov:1993es,DiMarco:2019czi,DiMarco:2021xzk} 
for perturbative reheating,
~\cite{Dolgov:1989us,Traschen:1990sw,Kofman:1994rk,Shtanov:1994ce,Kofman:1997yn,Greene:1997fu,Greene:1997ge,Greene:1998nh} 
for non-perturbative reheating,
and 
\cite{Bassett:2005xm,Frolov:2010sz,Allahverdi:2010xz,Amin:2014eta,Lozanov:2019jxc} 
for reviews).
 
This article reviews the fundamental aspects of the standard inflationary dynamics, focusing on the concept of the number of $e$-folds, 
i.e.,
the quantity that counts the number of exponential expansions of the universe, whose various definitions often give rise to some confusions in the literature.
In particular, Section~\ref{standard inflationary dynamics} is dedicated to the basics of the inflationary dynamic evolution in time, 
essentially described by a non-linear second-order differential equation for the homogeneous inflaton field.
Section~\ref{hamilton-jacobi inflationary dynamics} presents a derivation of the alternative Hamilton--Jacobi description of inflation, founded on a system of first-order differential equations.
Section~\ref{slow-roll parameters, inflationary perturbations and observables} introduces the hierarchies of slow-roll parameters, useful 
both for a reliable description of the slow-roll phase and 
for a simpler treatment of inflationary observables $\mathcal{O}(\phi)$ at some desired perturbative order.
Section~\ref{definitions of the number of efolds} contains a discussion of the general concept of the number of $e$-foldings ($e$-folds), $N_e$, together with 
a derivation of the related notions of
``total inflationary number of $e$-folds'', $N_T$, and of the fundamental ``number of $e$-folds before the end of inflation'', $N$.
Section~\ref{inflaton evolution and number of efolds} provides a detailed and exhaustive derivation of the inflaton equation for the case in which
the value of $N$ plays the role of a fundamental time-like variable.
Section~\ref{estimation of the number of e-folds before the end of inflation} discusses the computation of the number of $e$-folds before the end of inflation related to the observable cosmological scales.
In Section~\ref{inflationary observables and number of efolds}, 
solutions $\phi(N)$ of the inflaton equations are used to compute  numerical estimates of the main inflationary observables $\mathcal{O}(\phi)$ in terms of $N$. In particular, this recipe is applied to some representative classes of inflationary potentials, like monomial potentials, $\alpha$-attractor models and a class of non-linear Einstein--Cartan gravities, classically equivalent to models with a pseudoscalar slow-rolling inflaton. 
Finally, Section~\ref{summary and conclusions} provides conclusions and discussions.

In this manuscript, the particle natural units $\hbar=c=1$ and the ``mostly minus'' Lorentzian metric signature $(+,-,-,-)$ are used.
In addition, $M_p = \sqrt{1/8\pi G_N}$ denotes the reduced Planck mass, 
where $G_N$ is the Newton's gravitational constant.

\section{Standard Slow-Roll Inflationary Dynamics}
\label{standard inflationary dynamics}

The cosmological action for the slow-roll inflation can be written by adding to the 
classical Einsten--Hilbert action, describing the gravitational sector, a matter term whose particle content includes the inflaton (typically a scalar field),
together with other SM or BSM constituents.
However, on the one hand, the inflaton is expected to dominate over the remaining matter components.
On the other hand,  
as the inflationary evolution proceeds, the contributions due to additional matter degrees of freedom tend to become rapidly and increasingly negligible.
Therefore, a good description is given by
\begin{eqnarray}\label{totalact} 
    \mathcal{S}\left[g_{\mu\nu},\phi\right]&\sim& S_{\text{EH}} + S_{\phi} =  \nonumber\\
    &=&\int d^4x \sqrt{-g} \left( -\frac{M^2_p}{2} R + \frac{1}{2}\partial_{\mu}\phi\partial^{\mu}\phi - V(\phi)  \right),
\end{eqnarray}
where $\phi$ is the field (minimally coupled to gravity inflaton)
equipped with the (slow-roll) potential $V(\phi)$;
$R$ is the Ricci scalar curvature; and $g$ is the determinant of the metric tensor $g_{\mu\nu}$, 
assumed to be the one giving the line element:
\begin{equation}
\label{frw_metric}
    ds^2 = dt^2 - a(t)^2 dl^2, \quad dl^2 = \left[dr^2 + r^2\left(d\theta^2 + \sin^2\theta d\varphi^2 \right)\right].
\end{equation}
Here, $t$ represents the cosmic time, $a(t)$ is the (dimensionless) cosmic scale factor while  $dl^2$ is 
the (flat) euclidean line element of the metric describing three-dimensional spatial hypersurfaces at constant time in terms of the triplet of comoving coordinates $(r,\theta,\varphi)$.  
The validity of the assumption in Equation~\eqref{frw_metric} is supported by the fact that, 
during the slow-roll phase, the {hypothethical} spatial curvature contribution quickly becomes negligible.
The variation of the action of Equation~\eqref{totalact} with respect to $\phi$ provides the Klein--Gordon-like equation for the inflaton field, 
while the variation with respect to the gravitational metric field  $g_{\mu\nu}$ provides the usual Einstein field equations. 
The metric Ansatz in Equation~\eqref{frw_metric} determines a couple of 
non-linear differential equations for the cosmic scale factor $a(t)$ or for the Hubble rate $H=\dot{a}/a$, 
sourced, as usual, by the inflaton energy--momentum tensor $T_{\mu\nu}$
interpretable as that of a perfect fluid
with a given energy density $\rho_{\phi}$ and pressure $p_{\phi}$.
Indeed, the standard Einstein--Friedmann equations come out to be
\begin{equation}
\label{friedmann_equation_1}
    \left(\frac{\dot{a}}{a}\right)^2 = \frac{1}{3M^2_p}\rho_{\phi}
\end{equation}
and 
\begin{equation}
\label{friedmann_equation_2}
    \left(\frac{\ddot{a}}{a}\right) = -\frac{1}{6M^2_p}\left(\rho_{\phi} + 3p_{\phi}\right) ,
\end{equation}
where energy density and pressure can be identified with 
\begin{equation}
\label{inflaton_properties}
    \rho_{\phi} = \frac{1}{2}\dot{\phi}^2 + V(\phi), \quad p_{\phi} = \frac{1}{2}\dot{\phi}^2 - V(\phi) .
\end{equation}
The scalar wave equation for the inflaton reads 
\begin{equation}
\label{inflaton_equation_1}
    \ddot{\phi} + 3H\dot{\phi} + \frac{dV}{d\phi} = 0 ,
\end{equation}
and it is important to stress that the ``Hubble friction term''
containing the first-time derivative of the field is distinctive of the expansion of the universe.

The Einstein--Friedmann--Klein--Gordon (EFKG) system of Equations~\eqref{friedmann_equation_1}--\eqref{inflaton_equation_1} represents the standard tool in order to describe 
an (hypothetical) inflationary phase.
In addition, the system allows us to identify the boundary conditions providing 
 a suited accelerated inflating evolution of the universe.
In particular, Equation~\eqref{friedmann_equation_2} shows that the necessary requirement for a (positive) accelerated expanding phase
is the so-called \textit{first slow-roll condition}
~\cite{Steinhardt:1984jj,Liddle:1992wi,Liddle:1994dx}:
\begin{equation}
\label{first_slowroll_condition}
    \dot{\phi}^2\ll V(\phi).
\end{equation}
Moreover, the accelerated expansion must be sustained for a sufficiently long period in order to obtain a proper smooth and flat universe, 
in agreement with observations.
This is possible only if the inertial term in the inflaton Equation~\eqref{inflaton_equation_1} is negligible compared to the Hubble friction term
and the potential term.
In light of this, one has to impose a \textit{second slow-roll condition:}
\begin{equation}
\label{second_slowroll_condition}
    |\ddot{\phi}|\ll |3H\dot{\phi}|, \quad \Big|\ddot{\phi}\Big|\ll \Big|\frac{dV(\phi)}{d\phi}\Big| .
\end{equation}
The two slow-roll conditions can be translated into two corresponding \textit{slow-roll parameters:}
\begin{equation}
\label{original_slowroll_parameters}
    \epsilon(t) = -\frac{\dot{H}}{H^2}, \quad \eta(t) = -\frac{\ddot{\phi}}{H\dot{\phi}} .
\end{equation}
The first parameter can be derived just by noticing that $\ddot{a}/a = \dot{H} + H^2$, while the second term
is constructed by the ratio of the inertial and friction terms of the scalar wave Equation~\eqref{inflaton_equation_1}.
As mentioned, the smallness of the first one ($\epsilon\ll 1$) ensures the realization of the accelerated phase, 
while the smallness of the second one ($|\eta|\ll 1$) ensures that the accelerated phase lasts long enough to sufficiently stretch the universe in a way that is compatible with observations.
In general, the solutions of the first Friedmann equation and of the scalar field equation give a complete description
of the dynamics, i.e., determine the evolution of the scale factor $a(t)$ (or $H(t)$) and of the inflaton field $\phi(t)$.
The knowledge of the precise expansion rate of the universe is surely very interesting. 
However, the inflaton trajectory $\phi(t)$ represents the{ truly fundamental quantity}, 
so that
the system of equations is often reduced to the Equation~\eqref{inflaton_equation_1} coupled to 
\begin{equation}
    H^2 = \frac{1}{3M^2_p}\left( \frac{1}{2}\dot{\phi}^2 + V(\phi)  \right).
\end{equation}
They can be combined in a single non-linear second-order differential equation for the inflaton field
\begin{equation}
\label{inflaton_equation_2}
    \ddot{\phi} + \frac{\sqrt{3}}{M_p}\left( \frac{1}{2}\dot{\phi}^2 + V(\phi)  \right)^{1/2}\dot{\phi} + \frac{dV}{d\phi} = 0,
\end{equation}
promoted to a proper Cauchy problem by adding a reliable pair of initial conditions for the inflaton field and its first derivative at a given (initial) reference time $t_{i}$
\begin{equation}
\label{initial_conditions}
    \phi(t_i) = \phi_i, \quad \dot{\phi}(t_i) \ll 1 .
\end{equation}

The time $t_i$ can be arbitrary, although the most natural choice for it will be discussed later.
In principle, the solution of the Cauchy problem is obtained via numerical integration and provides an inflaton trajectory $\phi(t)$
as a function of the cosmic time. 
This function shows \textit{attractor-like behavior}~\cite{Belinsky:1985zd,Guo:2003zf,Urena-Lopez:2007zal,Remmen:2013eja} 
that is compatible with the inflationary phenomena 
and is extremely useful because it allows us to analyze the behavior of the inflaton field both 
\textit{(i)} during the fundamental slow-roll phase and \textit{(ii)} after the slow-roll phase, namely when the inflaton moves towards the true vacuum and begins to oscillate around it. 
Eventually, in the oscillating phase, the inflaton equation could be extended by adding phenomenological terms devoted to describe the decay of the inflaton modes into other particles. The related non-perturbative and perturbative reheating phases are fundamental to connect the inflationary epoch to 
the Friedmann--HBB dynamics in a way that preserves
the precise predictions of the Big Bang Nucleosynthesis~\cite{Bassett:2005xm,Frolov:2010sz,Allahverdi:2010xz,Amin:2014eta,Lozanov:2019jxc}.

\section{Hamilton--Jacobi Inflationary Dynamics}
\label{hamilton-jacobi inflationary dynamics}

The standard slow-roll dynamics is founded on a second-order scalar wave equation with the time-like variable given by the usual cosmic time $t$.
However, this scenario is not unique and other possibilities exist.
For example, an Hamilton--Jacobi
treatment of the inflationary equations based on a system of first-order (non-linear) differential equations in the time-domain has been introduced in 
\cite{Ellis:1990wsa}, while a more convenient version with the scalar field itself as the main time-like variable has been originally proposed in
\cite{Lidsey:1991zp}.
The starting point is the system of Friedmann equations that can be written just in terms of the Hubble rate:
\begin{equation}
\label{hubble_friedmann_equation_1}
    H^2 = \frac{1}{3M^2_p}\rho_{\phi}, 
\end{equation}
\begin{equation}
\label{hubble_friedmann_equation_2}
    \dot{H} + H^2 =-\frac{1}{6M^2_p}\left(\rho_{\phi} + 3p_{\phi}\right).
\end{equation}
Using the expressions of the energy density and pressure in Equation~\eqref{inflaton_properties}, the two Equations \eqref{hubble_friedmann_equation_1} and \eqref{hubble_friedmann_equation_2} can be rephrased in terms of the inflaton field and combined together in the single
non-linear equation
\begin{equation}
\label{inflaton_hubble_relation_1}
    \dot{\phi}^2 = -2M^2_p \dot{H} ,
\end{equation}
which relates the time derivative of the Hubble rate to the square of the inflaton time derivative.  It should be noticed that substituting $\dot{H}$ from  Equation~\eqref{inflaton_hubble_relation_1} inside the time derivative of Equation~\eqref{hubble_friedmann_equation_1} returns the original Klein--Gordon equation of the inflaton field, thereby confirming, in a sense, the redundancy of the original system of Equations~\eqref{friedmann_equation_1}--\eqref{inflaton_equation_1}.
In addition, Equation~\eqref{inflaton_hubble_relation_1} also suggests the alternative definition 
\begin{equation}
\label{epsilon_v2}
    \epsilon(t) = -\frac{\dot{H}}{H^2} = \frac{\dot{\phi}^2}{2M^2_p H^2}
\end{equation}
of the first slow-roll parameter. Moreover, one can substitute Equation~\eqref{inflaton_hubble_relation_1} in  Equation~\eqref{hubble_friedmann_equation_2},  
obtaining a single non-linear first-order differential equation for the Hubble rate, directly in terms of the scalar field potential
\begin{equation}
\label{hamilton_jacobi_time_1}
    3M^2_p H^2 + M^2_p\dot{H} = V(\phi) .
\end{equation}
Therefore, Equation~\eqref{inflaton_hubble_relation_1} and Equation~\eqref{hamilton_jacobi_time_1} determine a time-dependent first-order system of differential equations~\cite{Ellis:1990wsa}.
One should then trade the cosmic time variable for the scalar field itself to have a \textit{measure} of the Hubble variation in terms of the scalar field variation. 
It can be carried out by deriving the inflaton energy density $\rho_{\phi}$ with respect to time~\cite{Lidsey:1991zp}
and applying the Klein--Gordon equation to obtain
\begin{equation}
    \dot{\rho}_{\phi} = -3H\dot{\phi}^2.
\end{equation}
By dividing both members for $\dot{\phi}$, one obtains
\begin{equation}
    \frac{dt}{d\phi} = -\frac{3H}{\rho'_{\phi}} ,
\end{equation}
that implicitly returns the functional relation $t(\phi)$.
By using the derivative of Equation~\eqref{hubble_friedmann_equation_1} with respect to $\phi$, one can finally obtain the linear equation
\begin{equation}
\label{inflaton_hubble_relation_2}
    \dot{\phi} = -2M^2_p H'(\phi) .
\end{equation}
It should be noticed that the condition $\dot{\phi}\neq 0$ is crucial in this treatment.
Interestingly, the same result can also be achieved by using the simple relation
\begin{equation}
    \frac{dH}{dt} = \dot{\phi}\frac{dH}{d\phi}, 
\end{equation}
that always holds in the standard single-field slow-roll regime.
Equation~\eqref{inflaton_hubble_relation_2} provides two different scenarios.
The first corresponds to a decreasing scalar field ($\dot{\phi}<0$) during the slow-roll. 
In this case,  $H'(\phi)>0$ and, consequently, the Hubble rate also decreases with time, $\dot{H}(t)<0$.  
The second is the opposite one. 
An increasing scalar field ($\dot{\phi}>0$) during the slow-roll corresponds to $H'(\phi)<0$, resulting in an increasing-with-time Hubble rate, $\dot{H}(t)>0$.  Notice that a small variation in the time of the field does correspond to a small variation in the Hubble rate with the field, as heuristically expected. Equation~\eqref{inflaton_hubble_relation_2} can also be inserted into Equation~\eqref{hamilton_jacobi_time_1}, giving rise to the following 
additional equation in terms of derivative with respect to $\phi$:
\begin{equation}
\label{hamilton_jacobi_field_1}
    V(\phi) = 3M^2_pH^2(\phi) - 2M^4_pH'^2(\phi),
\end{equation}
The first-order Hamilton--Jacobi system of Equations \eqref{inflaton_hubble_relation_2} and \eqref{hamilton_jacobi_field_1}, 
must be decorated with a set of initial conditions for the Hubble rate and for the scalar field:
\begin{equation}
    H(\phi_i) = H_i, \quad \phi(t_i) = \phi_i .
\end{equation}

The Hamilton--Jacobi recipe furnishes an alternative window on the inflationary dynamics and has been used to study several features 
of the inflationary phase.
The first example is due to Salopek and Bondi~\cite{Salopek:1990jq}, 
who both discussed the analogy with the Hamilton--Jacobi formalism of classical mechanics and
the non-linear evolution of long-wavelength metric fluctuations in the context of stochastic inflation.
These equations are also very useful because they allow us to generate 
exact inflationary solutions starting from a chosen form of the $H(\phi)$ function, an aspect that has been intensively explored in~\cite{Muslimov:1990be,Barrow:1990vx,Barrow:1993hn,Barrow:1994nt,Barrow:1995xb}.

\section{Slow-Roll Parameters, Inflationary Perturbations, and Observables}
\label{slow-roll parameters, inflationary perturbations and observables}

The previously discussed Hamilton--Jacobi treatment of inflation represents a fundamental framework  
to introduce a complete hierarchy of field-dependent slow-roll parameters, as discussed by Liddle \textit{et al.}~\cite{Liddle:1994dx}.
In particular, it is possible to define the so-called Hubble slow-roll parameters (HSRPs) as
\begin{eqnarray}
\label{hsrp}
\epsilon_H(\phi)&=&2M^2_p\left(\frac{H'(\phi)}{H(\phi)} \right)^2, \\
\beta_H^{(n)}(\phi)&=&(2M^2_p)^n \left( \frac{(H')^{n-1} H^{(n+1)}}{H^n}  \right), \quad n\geq 1 .
\end{eqnarray}
In the literature, some higher-order HSRPs are also indicated with different symbols, like $\eta_H(\phi)=\beta_H^{(1)}$, $\xi_H^2(\phi)=\beta_H^{(2)}$, $\sigma_H^3(\phi)=\beta_H^{(3)}$, and so on.
It is important to note that the choice $\dot{\phi}<0$ naturally implies $\sqrt{\epsilon_H(\phi)}>0$ (see Equation~\eqref{inflaton_hubble_relation_2}). 
These parameters describe the properties of the inflationary dynamics, while their hierarchical smallness contains the initial conditions required to obtain a slow-roll inflationary phase.   In particular, $\epsilon_H(\phi)\ll 1$ and $|\eta_H(\phi)|\ll 1$ correctly reproduce the slow-roll conditions of Section~\ref{standard inflationary dynamics}. This is not a surprise since, for instance, the $\epsilon_H$ parameter in Equation~\eqref{hsrp} can be obtained by simply combining Equation~\eqref{epsilon_v2} with Equation~\eqref{inflaton_hubble_relation_2}. Needless to say, one can perform the same in terms of the potential, 
and indeed, one can define a second hierarchy, 
known as potential slow-roll parameters (PRSPs), as
\begin{eqnarray}
\label{psrp}
    \epsilon_V(\phi) &=& \frac{M^2_p}{2}\left(\frac{V'(\phi)}{V(\phi)}\right)^2, \\
    \beta^{(n)}_V(\phi) &=& (M^2_p)^n\left(\frac{(V')^{n-1}V^{(n+1)}}{V^n}\right) , \quad n\geq 1 .
\end{eqnarray}
Again, the higher-order PSRPs are also denoted as $\eta_V(\phi)$=$\beta^{(1)}$, $\xi_V^2(\phi)=\beta_V^{(2)}$, $\sigma_V^3(\phi)=\beta_V^{(3)}$,
and so on.   However, this set of parameters does not include the necessary conditions for having a slow-roll phase. Rather, it only classifies the flatness of the potential. In principle, the PSRPs are equivalent to the HSRPs if one were to add a constraint, i.e., the condition that the trajectory $\phi(t)$ approaches an \textit{attractor solution}~\cite{Liddle:1994dx}, corresponding to an asymptotic behavior given by the approximation
of the scalar wave equation $\phi \simeq - V'/ (3 H)$.
However, the HSRP and PSRP hierarchies are closely  correlated rather than independent.  Indeed, the Hubble parameters can always be expanded (up to an arbitrary order) in series of the potential parameters.  For example, 
the second-order expansions of the first three HSRPs are as follows:
\begin{eqnarray}
    \label{hsrp to psrp}
    \epsilon_H &=& \epsilon_V - \frac{4}{3}\epsilon^2_V + \frac{2}{3}\epsilon_V\eta_V + o(\beta^3_V),\\
    \eta_H &=& \eta_V - \epsilon_V + \frac{8}{3}\epsilon^2_V + \frac{1}{3}\eta^2_V - \frac{8}{3}\epsilon_V\eta_V + o(\beta^3_V) ,\\
    \xi^2_H &=& \xi^2_V - 3\epsilon_V\eta_V + 3\epsilon^2_V + o(\beta^3_V).
\end{eqnarray}
Vice versa, the PSRPs admit an exact expression in terms of the HSRPs (see~\cite{Liddle:1994dx} for details), 
as one can deduce from Equation~\eqref{hamilton_jacobi_field_1}.
If needed, one can as well expand the PSRPs in a series of HSRPs.
The inflationary stage is notoriously characterized by a decrease in the comoving Hubble horizon (or radius) $1/a H$
and by an ubiquitous production of vacuum fluctuations on all comoving subhorizon scales $k\gg a H$
of scalar fields with an effective mass smaller than the Hubble rate.
In the simplest case, {this involves only the inflaton field driving the expansion.} 
These inflaton fluctuations naturally turn into metric fluctuations around the homogeneous FLRW spacetime~\cite{Riotto:2002yw},
resulting in a metric tensor of the form
\begin{equation}
\label{metric_fluctuations}
    g_{\mu\nu}\sim g_{\mu\nu}^{\text{FLRW}} + \delta g_{\mu\nu}^{(1)} + \delta g_{\mu\nu}^{(2)} + ...
\end{equation}
In light of this, one can single out
three main types of fluctuation modes, i.e., scalar (S), vector (V), and tensor (T) modes.
In the first approximation, one can consider \textit{small} fluctuations around the FLRW metric
and simply take into account the linear contributions of the expansion in Equation~\eqref{metric_fluctuations}~\cite{Riotto:2002yw}.
The first-order approach provides simple field equations for the three types of modes
in the space of comoving $k$. {Such modes result in a complete decoupling from each other at a given scale $k$.} 
In particular, the solution of the field equations suggests that 
the vector modes can be neglected, while 
scalar and tensor modes are naturally stretched beyond the (slowly changing) inflationary Hubble horizon 
by the accelerated expansion, quickly freezing to superhorizon classical metric perturbations $k\ll aH$ at a conserved value.
The related period is usually called \textit{horizon crossing} and can be referred to \textit{any} perturbed arbitrary scale $k$.
However, for the sake of convenience, it is better to reserve
such a definition for the indication of the specific crossing of an \textit{observable} perturbed scale, 
or the period in which the entire package of (human) \textit{observables} perturbed scales are stretched out.

{The superhorizon solutions for the amplitude of scalar and tensor modes can be used to construct the \textit{so-called} 
power spectra of perturbations, that can be written as}
\begin{equation}
\label{power_spectra}
    P_S(k) \sim \frac{1}{8\pi^2 M^2_p}\frac{H^2}{\epsilon_H}\Bigl|_{k=aH} \quad ; \quad \quad
    P_T(k) \sim \frac{2}{\pi^2}\frac{H^2}{M^2_p}\Bigl|_{k=aH} \ ,
\end{equation}
where only $P_S$ happens to depend upon $\epsilon_H$, while $P_T$ does not.
It is important to stress that the solutions of Equations~\eqref{power_spectra} do not represent the power spectra \textit{at} horizon crossing.
Indeed, they are the asymptotic solutions only \textit{written} in terms 
of the horizon crossing phase $k\sim a H$ (for details, see~\cite{Lidsey:1995np}).
Moreover, since the evolution of the mode(s) $k$ depends on the scalar field, Equations~\eqref{power_spectra}
can be thought of as functions of $\phi$ at horizon crossing, via $H$ and $\epsilon_H$.
{
The solutions of Equations~\eqref{power_spectra} also allow us to conclude that single-field slow-roll inflation generates 
almost-scale-invariant metric perturbations with a corresponding Gaussian probability density function (PDF)~\cite{Riotto:2002yw}.}

The power spectra also allow us to derive the so-called inflationary observables, $\mathcal{O}(\phi)$,
that inherit the field dependence~\cite{Lidsey:1995np}
\begin{eqnarray}
\label{spectra_parameters_hsrp}
    n_S(\phi) &\sim& 1 - 4\epsilon_H(\phi) + 2\eta_H(\phi) ,\\
    n_T(\phi) &\sim& -2\epsilon_H(\phi), \quad r(\phi) \sim 16\epsilon_H(\phi) ,\\
    \alpha_S(\phi) &\sim& -2\xi^2_H(\phi) + 10\epsilon_H(\phi)\eta_H(\phi) - 8\epsilon^2_H(\phi),\\
    \alpha_T(\phi) &\sim& -4\epsilon_H(\phi)(\epsilon_H(\phi)-\eta_H(\phi)) ,
\end{eqnarray}
where $n_S$ is the scalar tilt, $n_T$ is the tensor tilt, $r$ is the tensor-to-scalar ratio, $\alpha_S$ is the running of the scalar tilt, and $\alpha_T$ represents the running of tensors. 
Currently, the parameter $n_T$ holds more theoretical than practical interest 
due to the lack of a (direct or indirect) detection of an inflationary gravitational wave background.
These functions describe how the value of the power spectra in Equation~\eqref{power_spectra} changes with (the log of) $k$~\cite{Lidsey:1995np}.
Typically, the inflationary stage predicts tiny inflationary observables
{confirming the almost-scale-invariant behavior in the $k$ space for the power spectra of Equation~\eqref{power_spectra}.}
The specific dependency of the inflationary observables from the scalar field value $\phi$ is not simple to derive,
because it needs the knowledge of the form of the Hubble rate $H(\phi)$, which is strongly model-dependent.
Fortunately, it is possible to use the expansions of the HSRPs in terms of the PSRPs
to convert the $H(\phi)$ dependency to a potential one, obtaining~\cite{Lidsey:1995np}:
\begin{eqnarray}
\label{spectra_parameters_psrp}
    n_S(\phi) &\sim& 1 - 6\epsilon_V(\phi) + 2\eta_V(\phi), \\
    n_T(\phi) &\sim& -2\epsilon_V(\phi), \quad r(\phi) \sim 16\epsilon_V(\phi), \\
    \alpha_S(\phi) &\sim& -2\xi^2_V(\phi) + 16\epsilon_V(\phi)\eta_V(\phi) - 24\epsilon^2_V(\phi),\\
    \alpha_T(\phi) &\sim& -4\epsilon_V(\phi)(2\epsilon_V(\phi)-\eta_V(\phi)).
\end{eqnarray}
In this way, the inflationary observables are explicitly determined by (the flatness of) the scalar potential,
which are  functions of $\phi$ through $V(\phi)$.
The cosmological observations reveal that the amplitude of the scalar perturbations is of order
$P_S\sim 2\times 10^{-9}$~\cite{COBE:1992syq}, the scalar spectral index, measured at the reference comoving scale $k\sim 0.002$ Mpc$^{-1}$, is
$n_S = 0.9649\pm0.0042$ at $68\%$ of confidence level  (CL)~\cite{Planck:2018jri}, 
while the BICEP/Keck upper limit on the tensor-to-scalar ratio at the reference comoving scale $k\sim 0.05$ Mpc$^{-1}$, 
results  $r < 0.036$ at $95\%$ CL~\cite{BICEP:2021xfz}.

The observables above enable testing a single-field slow-roll inflation model with potential $V(\phi)$ against current observations by varying the field as a parameter. This can be performed by selecting a \textit{reasonable} sample of scalar field values from the potential plateau. However, this method is approximate, as the chosen field values may not correspond to those at horizon crossing for the relevant scale(s) $k$. A reliable inflaton trajectory is needed, which requires solving the inflaton equation of motion \eqref{inflaton_equation_2} with proper initial conditions (Equation~\eqref{initial_conditions})
for (the horizon crossing of) the scales of interest, such as $0.002$ Mpc$^{-1}$.

This task is rather delicate because it is not easy to select (specific) time(s) and durations of the desired horizon crossing. 
Therefore, the introduction of a new \textit{time-like variable} is necessary. 
The following three sections will be devoted to presenting this variable, exploring its utility, and discussing its natural values
in the context of horizon crossing.

\section{Definitions of the Number of \boldmath{$e$}-Folds}
\label{definitions of the number of efolds}

In inflationary cosmology, the evolution of the {inflating universe's patch} 
can be described, to a first approximation, by an explicit exponential term.
Indeed, by definition, the Hubble rate expression can be interpreted as an ordinary differential equation once 
equipped with an initial condition for the cosmic scale factor at a generic {slow-roll} trigger time $t_i$ (see Equation~\eqref{initial_conditions}):
\begin{equation}
    \frac{\dot{a}}{a} = H(t), \quad a(t_i) = a_i ,
\end{equation}
whose solution obviously reads
\begin{equation}
    a(t) = a_i e^{\int_{t_i}^{t} d\tau \, H(\tau)} \quad  \quad (t>t_i) .
\end{equation}
Imposing the first slow-roll condition in Equation~\eqref{first_slowroll_condition}, 
 the Hubble function depends only on the value acquired by the inflaton (potential) energy density during the \mbox{inflationary epoch}
\begin{equation}
\label{hubble_scale}
    H^2(\phi)\sim \frac{1}{3M^2_p}\rho_{\phi}, \quad \rho_{\phi}\sim V(\phi) ,
\end{equation}
corresponding to a ``transient'' ideal fluid with an equation of state (EoS) $w_{\phi} = -\rho_{\phi}/p_{\phi}\sim -1$.
In general, the time integral of the Hubble rate is called the \textit{number of $e$-folds} 
of the expansion from $t_i$ to the generic cosmic time $t$. 
It measures exactly the rate of exponential growing of the universe during inflation, and can be considered as a function of the \mbox{upper extreme}:
\begin{equation}
\label{number_of_efolds}
    N_e(t) = \int_{t_i}^{t} d\tau \, H(\tau), \quad \frac{dN_e}{dt} = H(t).
\end{equation}
If one assumes $V(\phi)$ as an almost constant plateau (for a reasonable range of the scalar field values) of height $\sim M^4_{\text{inf}}$,
the result of the integration is a linear function of time $\sim H_{\text{inf}}(t-t_i)$, with $H_{\text{inf}}\sim M^2_{\text{inf}}/M_p$.
Generically, the integral is more complicated, but still, $N(t)$ is an \textit{increasing} function of time, with 
 a positive derivative also causing the cosmic scale factor $a(t)$ to monotonically increase.
Equation~\eqref{number_of_efolds} allows us to write down the total number of $e$-folds of an inflationary period from $t_i$ to a final time $t_{\text{end}}$ as
\begin{equation}
    N_T = \int_{t_{i}}^{t_{\text{end}}} d\tau \, H(\tau).
\end{equation}

If {single-field slow-roll} inflation is assumed to start at a time $t_i$ immediately following a primordial quantum gravity phase 
and thought to propagate below an appropriate BSM phase
(for instance, a GUT phase), the total number of exponential expansions could amount to millions---an
information that may be indicative of the real size of the causal-connected universe, the so-called \textit{particle horizon}
\cite{Kolb:1990vq,Mukhanov:2005sc,Weinberg:2008zzc,Gorbunov:2011zzc}.
However, it is commonly believed
that the observable perturbed scales (and in some sense, the current observable universe)
has been generated at certain times during the last stages of the {inflationary expansion}.
In light of this, one can introduce a new convenient variable, the \textit{number of $e$-folds before the end of inflation} 
\begin{equation}
\label{deltanumbofef}
    \Delta N(t,t_{end}) = N_T - N_e(t) = \int_{t_i}^{t_{end}} d\tau \, H(\tau) - \int_{t_i}^{t} d\tau \, H(\tau) =\int_{t}^{t_{end}} d\tau \, H(\tau),
\end{equation}
obtained by subtracting {from} 
$N_T$ the number of $e$-folds that precede the beginning of the ``interesting phase'',
where the observable perturbed scales are generated.
This variable is decreasing in time, since
\begin{equation}
    \frac{d \Delta N(t,t_{end})}{dt} = - \frac{dN_e}{dt} < 0 ,
\end{equation}
and can be used as a \textit{backwords}
time-like variable, vanishing when $t=t_{\text{end}}$.

The number of $e$-folds before the end of inflation is often rewritten in terms of the inflaton field  $\phi$ along the flat potential
direction and of the first HSRP $\epsilon_H(\phi)$.
In particular, formally inverting the time profile of $\phi$, one can turn the time integral into a field-space integral, obtaining 
\begin{equation}
    \Delta N(\phi,\phi_{end}) = \int_{\phi}^{\phi_{end}} d\widetilde{\phi} \ \frac{H(\widetilde{\phi})}{\dot{\widetilde{\phi}}}.
\end{equation}
Using Equation~\eqref{inflaton_hubble_relation_2}, assuming $\dot{\phi}<0$ and inserting the parameter $\epsilon_H(\phi)$, one finally obtains
\begin{equation}
\label{number_of_efolds_before_end_of_inflation}
    \Delta N(\phi,\phi_{end}) = \frac{1}{M_p} \int_{\phi_{end}}^{\phi} d\widetilde{\phi}  \ \frac{1}{\sqrt{2\epsilon_H(\widetilde{\phi})}}.
\end{equation}
In the following, we define
\begin{equation}
\label{number_of_efolds_before_end_of_inflation_final}
    N(\phi) \equiv \Delta N(\phi, \phi_{\text{end}}).
\end{equation}
The reader should be aware that many authors label the same quantity with other notations, e.g., $N_e$ or $N_*$.

It should be noticed that
Equation~\eqref{number_of_efolds_before_end_of_inflation}, 
under the approximation $\epsilon_H(\phi)\sim \epsilon_V(\phi)$, 
is equivalent to the inverse solution of the first-order differential equation
\begin{equation}
    \frac{d\phi}{dN} \sim M_p \sqrt{2\epsilon_V(\phi)}
\end{equation}
that provides
the scalar field profile as a function of $N$, $\phi(N)$.
As aforementioned, such field trajectory is not an exact solution, as it is rather the slow-roll limit of a more general equation, and is to be discussed in the next section.

\section{Inflaton Evolution and Number of $e$-Folds}
\label{inflaton evolution and number of efolds}

In Section~\ref{standard inflationary dynamics}, the system of Einstein--Friedmann--Klein--Gordon equations was introduced, which provides the function $a(t)$ and the global inflaton behavior $\phi(t)$.
Nevertheless, as anticipated at the end of Section~\ref{slow-roll parameters, inflationary perturbations and observables},
the knowledge of the time profile of the field $\phi$ is not very useful 
for the numerical estimate of the inflationary observables $\mathcal{O}(\phi)$. 
These quantities are closely related to the horizon crossing window, so it would be advisable to compute them for a proper range of the number of $e$-folds $N$.
In this respect, it would be much better to obtain inflaton solutions as explicit functions of the number of $e$-folds
\begin{equation}
    \mbox{ Observable } \mathcal{O}[\phi(t)] \quad\rightarrow\quad \mbox{ Horizon Crossing }\quad\rightarrow\quad N \quad\rightarrow\quad \mathcal{O}[\phi(N)],
\end{equation}
in such a way that it is possible to literally sample $\mathcal{O}(\phi)$ in terms of $N$.  To this aim, it is useful to rewrite the Klein--Gordon equation with $N$ (in place of the cosmic time) as the basic time-like  variable.  
Its derivation, often overlooked in the current literature (see, however,~\cite{Martin:2013tda} for a brief description) is now reviewed in detail. 

The relation between time and $N$ is contained in the equation $\dot N = - H$, which results from the combination
of Equation~\eqref{number_of_efolds} and Equation~\eqref{deltanumbofef}.

Given a generic function whose time dependence  $f(t) = f[N(t)]$ is through $N$, it is straightforward to derive the expressions for the first and second derivatives in the form
\begin{align}
    \frac{df}{dt} &= - H\frac{df}{dN} ,\\ \nonumber\\
    \frac{d^2f}{dt^2} &= H^2(N) \, \epsilon(N) \, \frac{df}{dN} + H^2(N) \, \frac{d^2f}{dN^2} ,
\end{align}
where the relation  
\begin{equation}
\label{epsilon_n}
    \epsilon(N)= \epsilon(t(N)) = -\frac{\dot{H}(t)}{H^2} = \frac{1}{H}\frac{dH}{dN}
\end{equation}
has been used to insert  the first slow-roll parameter $\epsilon(N)$. 
Applying the above relations to $\phi(N)$ inside the Klein--Gordon equation, one obtains
\begin{equation}
    H^2(N)\epsilon(N)\frac{d\phi}{dN} + H^2(N)\frac{d^2\phi}{dN^2} - 3H^2(N) \frac{d\phi}{dN} + \frac{dV}{d\phi} = 0 ,
\end{equation}
or
\begin{equation}
    \frac{d^2\phi}{dN^2} + \epsilon(N)\frac{d\phi}{dN} - 3\frac{d\phi}{dN} + \frac{1}{H^2(N)}\frac{dV}{d\phi} = 0.
\end{equation}
It is useful to highlight that  Equation~\eqref{friedmann_equation_1} gives back $H$ as a function of $N$ in the form
\begin{equation}
    H^2(N) = \frac{2V(\phi)}{ 6M^2_p - \left(\frac{d\phi}{dN}\right)^2  } ,
\end{equation}
while Equation~\eqref{inflaton_hubble_relation_1} in terms of $N$ provides 
\begin{equation}\label{slowrollepsenne}
    \left(\frac{d\phi}{dN}\right)^2 = 2M^2_p\epsilon(N). \quad 
\end{equation}
As a result, the Hubble rate can be written as
\begin{equation}
    H^2(N) = \frac{1}{M^2_p}\frac{V(\phi)}{3 - \epsilon(N)}
\end{equation}
in terms of the potential and the ``new'' slow-roll parameter $\epsilon(N)$, introduced in Equation~\eqref{epsilon_n}.
Performing an inversion of sign $N\rightarrow -N$ in order to allow for a forward integration over the time-like variable, the Cauchy problem related to the Klein--Gordon equation finally reads 
\begin{align}
\label{inflaton_equation_efolds}
    &\frac{d^2\phi}{dN^2} + (3-\epsilon(N))\frac{d\phi}{dN} + (3-\epsilon(N))M_p\sqrt{2\epsilon_V(\phi)} = 0, \nonumber\\
    &\phi(N_i) = \phi_i, \quad\quad \dot{\phi}_i\ll1.
\end{align}
The system is a second-order differential problem. It is non-linear because $\epsilon(N)$ contains a derivative squared. Moreover, the first derivative term plays the same role as the Hubble friction in the original formulation.  
The Klein--Gordon equation needs to be numerically integrated  to provide the exact inflaton trajectory in terms of the  $e$-folds number.
However, the slow-roll limit corresponding to the conditions
\begin{equation}
    \epsilon(N)\ll 1, \quad\quad \frac{d^2\phi}{dN^2}\ll\left(\frac{d\phi}{dN},\ M_p\sqrt{2\epsilon_V(\phi)}\right),
\end{equation}
can be used to reduce the equation to the linear one
\begin{equation}
    \frac{d\phi}{dN} + M_p\sqrt{2\epsilon_V(\phi)} \sim 0 ,
\end{equation}
that simply matches the equation met in the previous section (with $N\rightarrow -N$). 
Of course, the integration of the linear equation correctly reproduces the (first-order) definition of the
number of $e$-folds before the end of inflation
\begin{equation}
\label{inflaton_efolds_approx}
   N(\phi) \equiv \Delta N(\phi,\phi_{end}) \sim \frac{1}{M_p} \int_{\phi_{end}}^{\phi} d\widetilde{\phi} \frac{1}{\sqrt{2\epsilon_V(\widetilde{\phi})}}, 
\end{equation}
where the inflaton value $\phi_{\text{end}}$ is computed from the corresponding slow-roll breaking condition
\begin{equation}
    \epsilon_V(\phi_{end}) \sim 1.
\end{equation}

This treatment suggests that, given a scalar potential $V(\phi)$, an approximate analytic solution $\phi(N)$ can be simply obtained by inverting Equation~\eqref{inflaton_efolds_approx}, if possible.
Such a solution allows us to choose an initial $N$ and then to compute a related initial field value
for the Cauchy problem in Equation~\eqref{inflaton_equation_efolds}.

{The numerical integration and the related field function should be given over a reliable range of $N$
associated with the horizon crossing of the reference-observable scales.}
In principle, nothing prevents assuming arbitrary initial conditions for the slow-rolling field. However, it is certainly better to consider simplified conditions already compatible with the horizon crossing of the observable perturbed scales.
In the next section, we present a well-codified strategy to select the value of $N$ related exactly to the epoch in which the relevant observable scales cross the inflationary Hubble horizon.

\section{Estimates of the Number of \boldmath{$e$}-Folds before the End of Inflation}
\label{estimation of the number of e-folds before the end of inflation}

The number of $e$-folds occurring before
the end inflation, $N$, can be estimated in a very simple way in terms of generic properties of the inflationary phase 
and the features qualifying the postinflationary evolution.  
In this context, it is interesting to consider a comoving scale $k$ that crosses the Hubble horizon at a certain time $t$ during inflation, $k\sim a H$, and the comoving scale corresponding to the size of the current Hubble horizon, $k_0\sim a_0 H_0$.
The ratio of the two comoving scales
\begin{equation}
     \frac{a H}{a_0 H_0} = \frac{k}{k_0} = \frac{\lambda_0}{\lambda} 
\end{equation}
is naturally constant throughout the evolution of the universe.
This ratio can also be \mbox{split as}
\begin{equation}
    \frac{k}{a_0 H_0} = \frac{a}{a_{\text{end}}}\frac{a_{\text{end}}}{a_{\text{reh}}}\frac{a_{\text{reh}}}{a_0}\frac{H}{H_0}, 
\end{equation}
where the first partial ratios are referred as the inflationary evolution, the reheating epoch, and the HBB epoch, respectively, while 
the last fraction represents the ratio between the Hubble scale at the inflationary time $t$ and the current one.
In order to obtain the number of $e$-folds, it is convenient to take the logarithm of the ratio, obtaining 
\begin{eqnarray}
    \log\frac{k}{a_0 H_0} &=& \log\left(\frac{a}{a_{\text{end}}}\right) + \log\left(\frac{a_{\text{end}}}{a_{\text{reh}}}\right) + 
    \log\left(\frac{a_{\text{reh}}}{a_0}\right) + \log\left(\frac{H}{H_0}\right) \\ \nonumber
                          &=& -N - N_{\text{reh}} - N_{\text{HBB}} + \log\left(\frac{H}{H_0}\right).
\end{eqnarray}
Here, as said, $N$ represents the number of $e$-folds occurring between the generic time $t$ and the end of inflation, $t_\text{end}$, 
while $N_{\text{reh}}$ describes the number of $e$-folds during the reheating phase
and $N_{\text{HBB}}$ indicates the $e$-folds related to the standard Big Bang phase.
One can thus write the following:
\begin{equation}
\label{minimal_efolds_1}
    N = -\log\frac{k}{a_0 H_0} - N_{\text{reh}} - N_{\text{HBB}} + \log\frac{H}{H_0} 
\end{equation}
Notice that the first contribution is just a numerical term that depends on the chosen comoving scale $k$.
$N_{\text{reh}}$, on the other hand, can be estimated by assuming a global and effective description of the reheating phase.
In particular, one can suppose a reheating phase characterized by a single, global fluid with an average equation of state (EoS) $w_{\text{reh}}$
and a conserved average energy density described by a proper Boltzmann equation:
\begin{equation}
\label{global_reheating_boltzmann_equation}
    \dot{\rho}_{\text{reh}} + 3H(1+w_{\text{reh}}, \sim 0 .
\end{equation}
The solution of Equation~\eqref{global_reheating_boltzmann_equation}
between $t_{\text{end}}$ and $t_{\text{reh}}$
allows us to express $N_{\text{reh}}$ in terms of the average $w_{\text{reh}}$, the energy density 
at the end of slow-roll inflation $\rho_{\text{end}}$, and the energy density at the end of the reheating $\rho_{\text{reh}}$:
\begin{equation}
\label{reheating_efolds}
    N_{\text{reh}} \sim \frac{1}{3(1+w_{\text{reh}})}\log\frac{\rho_{\text{end}}}{\rho_\text{reh}} .
\end{equation}
The third contribution in Equation~\eqref{minimal_efolds_1} can be easily extracted from the fundamental definition of an HBB cosmology, i.e.,
the conservation of the comoving entropy density:
\begin{equation}
    g_S(t_{\text{reh}})a^3_{\text{reh}}T^3_{\text{reh}} = g_S(t_0) a^3_0 T^3_0 , 
\end{equation}
where $g_S$ is the number of the relativistic degrees of freedom contributing to the entropy density,
$T_{\text{reh}}$ is the temperature of the relativistic matter close to the reheating scale,
and $T_0$ is the current temperature of the relativistic component
of the universe, mainly CMB photons.
Remembering that 
\begin{equation}
    \rho_{\text{reh}}\sim \frac{\pi^2}{30}g_E(t_{\text{reh}})T^4_{\text{reh}} ,
\end{equation}
where $g_S$ is the number of the relativistic degrees of freedom contributing to the energy density,
and by assuming that $g_S(t_{\text{reh}})\sim g_E(t_{\text{reh}})\sim g_{\text{reh}}$, one can infer that
\begin{equation}
\label{hbb_efolds}
    N_{\text{HBB}}\sim \log T_0 - \frac{1}{4}\log\rho_{\text{reh}} -
     \frac{1}{12} \ln g_{reh} + 
    \log\left(\frac{43}{11}\right)^{\frac{1}{3}}\left(\frac{\pi^2}{30}\right)^{\frac{1}{4}} .
\end{equation}
The substitution of the relations Equation~\eqref{reheating_efolds}, Equation~\eqref{hbb_efolds}, and Equation~\eqref{hubble_scale} into Equation~\eqref{minimal_efolds_1}
leads to
\begin{equation}
\label{efolds_estimation_1}
    N \sim \zeta -  
    \frac{1}{3(1+w_{\text{reh}})}\ln\left(\frac{\rho_{\text{end}}}{\rho_\text{reh}}\right) + 
    \frac{1}{4}\ln\left(\frac{V^2}{M^4_p\rho_{\text{reh}}}\right)
\end{equation}
where
\begin{equation}
\zeta = -\ln\left(\frac{k}{a_0 H_0}\right) + \ln\left(\frac{T_0}{H_0}\right) + \zeta_0 \ 
\end{equation}
and $\zeta_0$ is just a numerical factor
\begin{equation}
\zeta_0 = - \frac{1}{12} \ln g_{reh} + \frac{1}{4}\ln\left(\frac{1}{9}\right) + 
\ln\left(\frac{43}{11}\right)^{\frac{1}{3}}\left(\frac{\pi^2}{30}\right)^{\frac{1}{4}} .
\end{equation}

Equation~\eqref{efolds_estimation_1} describes the number of $e$-folds 
related to the \textit{horizon exit} of the mode $k$ in terms of 
both the properties of the inflationary phase ($V(\phi)$, $\rho_{\text{end}}$)
and the properties of the postinflationary phase, relative to reheating and HBB cosmology epochs.
In particular, considering for simplicity
an almost instantaneous reheating phase and neglecting the subdominant
numerical term $\zeta_0$, one can obtain the following simplified expression for the number of $e$-folds:

\begin{equation}
\label{efolds_estimation_2}
    N\sim -\ln\left(\frac{k}{a_0 H_0}\right) + \ln\left(\frac{T_0}{H_0}\right) + \ln\frac{M_{\text{inf}}}{M_p} .
\end{equation}

It is important to emphasize that $N$ depends on the chosen ``pivot'' comoving wave number $k$.  It is worth mentioning that a smaller comoving scale $k<a_0 H_0$ would obviously give a smaller value of $N$, reflecting the fact that the smaller the comoving scale crossing the horizon, the lower the number of $e$-folds before the end of inflation. It is customary to focus on perturbed scales that can be experimentally probed, like that of the Planck mission, $k\sim 0.002$ Mpc$^{-1}$.
Remembering that $a_0=1$, $H_0\sim 1.75\times 10^{-42}$ GeV and $T_0=2.3\times 10^{-13}$ GeV, one obtains
\begin{equation}
    N \sim 64 + \ln\frac{M_{\text{inf}}}{M_p} .
\label{NinPlanckapp}\end{equation}

In Equation~\eqref{NinPlanckapp}, $N$ has a mild dependency on the inflationary scale.
It is indeed possible to demonstrate that it lies in the range $60$--$50$~\cite{Liddle:2003as,German:2022sjd} for $M_{\text{inf}}\sim (10^{16}$--$10^{13})$ GeV.
A more complicated postinflationary phase 
(see, for example,~\cite{Das:2015uwa,Cicoli:2016olq,Maharana:2017fui,DiMarco:2018bnw}) would induce modifications to the standard expression of Equation~\eqref{efolds_estimation_1}. 

{In the next section, the first-order computation of inflationary observables in terms of the variable $N$ 
for several paradigmatic classes of inflationary models is presented.
In addition, numerical predictions for the commonly chosen value of $N \sim 60$ and their compatibility with current experimental constraints
are also discussed.}

\section{Inflationary Observables and Number of \boldmath{$e$}-Folds}
\label{inflationary observables and number of efolds}

In the realm of the single-field slow-roll inflationary models,  
the simplest example is surely given by the class of  monomial potentials.
This class of potentials was firstly introduced contextually to the chaotic inflation scenario~\cite{Linde:1983gd}
and emerge, for instance, in a number of BSM-motivated cosmologies like those within supergravity~\cite{Kawasaki:2000yn,Takahashi:2010ky,Nakayama:2010kt,Nakayama:2010sk,Nakayama:2010ga,Harigaya:2012pg}
or in superstring theories (see, e.g., axion monodromy inflation~\cite{Silverstein:2008sg,McAllister:2008hb,McAllister:2014mpa}).
The form of the potential is 
\begin{equation}
    V(\phi) = M^4_{\text{inf}}\left(\frac{\phi}{M_p}\right)^n,
\end{equation}
which is often also written as
\begin{equation}
    V(\phi) = \lambda_n\phi^n, \quad \lambda_n = M^4_{\text{inf}}M_p^{-n} ,
\label{powerpotent}\end{equation}
where the parameter $\lambda_n$ is essentially the coupling constant of the inflaton self-interaction.
This class of models falls within the larger class of \textit{large-field inflation} 
since the vacuum expectation value of the inflaton field reduces from super-Planckian values $\phi\gg M_p$
to the vacuum value in $\phi=0$ with $\dot{\phi}<0$.
The most known scenarios are the quadratic case 
\begin{equation}
    V(\phi) = \frac{1}{2}m_{\phi}^2\phi^2, \quad m^2_{\phi} = \frac{2M^4_{\text{inf}}}{M^2_p} ,
\end{equation}
and the quartic case
\begin{equation}
    V(\phi)=\frac{\lambda_{\phi}}{4}\phi^4, \quad \lambda_{\phi} = 4\frac{M^4_{\text{inf}}}{M^4_p} ,
\end{equation}
originally discussed by Andrei Linde as the simplest examples of chaotic inflation~\cite{Linde:1983gd}.
The $k$-th derivative of the scalar potential in Equation~\eqref{powerpotent} 
\begin{equation}
V^{(k)}(\phi) =\frac{M^4_{\text{inf}}}{M_p^k} \, \frac{\Gamma(n+1)}{\Gamma(n-k+1)} \, \left(\frac{\phi}{M_p}\right)^{n-k} , \quad k\leq n ,
\end{equation}
allows us to write, for instance, the first three potential parameters:
\begin{eqnarray}
    \epsilon_V(\phi) &=& \frac{n^2}{2}\left(\frac{M_p}{\phi}\right)^2, \\
    \eta_V(\phi) &=& n(n-1)\left(\frac{M_p}{\phi}\right)^2, \\
    \xi^2_V(\phi) &=& n^2(n-1)(n-2)\left(\frac{M_p}{\phi}\right)^4 .
\end{eqnarray}
Thus, one can derive the expression of the scalar power spectrum as 
\begin{equation}
    P_S(\phi) \sim \frac{1}{12\pi^2n^2}\left(\frac{M_{\text{inf}}}{M_p}\right)^4\left(\frac{\phi}{M_p}\right)^{n+2}
\end{equation}
as well as the inflationary observables introduced in Section~\ref{slow-roll parameters, inflationary perturbations and observables} 
as explicit functions of the scalar field:
\begin{eqnarray}
\label{monomial_predictions_field}
    n_S(\phi) &\sim& 1 - n(n+2)\left(\frac{M_p}{\phi}\right)^2, \\
    r(\phi) &\sim& 8n^2\left(\frac{M_p}{\phi}\right)^2, \\ 
    \alpha_S(\phi) &\sim& -2n^2(n+2)\left(\frac{M_p}{\phi}\right)^4, \\
    \alpha_T(\phi) &\sim& 2n^3\left(\frac{M_p}{\phi}\right)^4 .
\end{eqnarray}

The condition $\dot{\phi}<0$ naturally suggests how the scalar spectral index tends to decrease while the tensor-to-scalar ratio
has an opposite behavior.
As discussed in Section~\ref{slow-roll parameters, inflationary perturbations and observables}, 
one can roughly deduce the value of the cosmological parameters
by considering a reasonable set of inflaton configurations 
 in  the slow-roll plateau region of the scalar potential $V(\phi)$.
Nevertheless, a preferred approach is to determine the inflaton values by solving the inflaton equation of motion \eqref{inflaton_equation_efolds}, 
derived to this aim in Section~\ref{inflaton evolution and number of efolds}.
The associated numerical solution gives an inflaton trajectory $\phi(N)$ that can be used 
to obtain a final, numerical evolution  of the interesting observables with respect to $N$.  Finally, first-order analytic estimates of the cosmological parameters in terms of $N$ can be very useful as well. To achieve them, 
one can consider the slow-roll limit of the exact Equation~\eqref{inflaton_efolds_approx} for $N$ 
\begin{equation}
    N(\phi)\sim \frac{1}{n M_p^2}\frac{1}{2}\left(\phi^2 - \phi_{end}^2 \right) , 
\end{equation}
and try to invert it explicitly. The result is 
\begin{equation}
    \left(\frac{\phi}{M_p}\right)^2 \sim 2nN + \left(\frac{\phi_{end}}{M_p}\right)^2, 
\end{equation}
where the final field value, obtained by the condition $\epsilon_V(\phi_{end})\sim 1$, comes out to be
\begin{equation}
     \left(\frac{\phi_{end}}{M_p}\right)^2 \sim \frac{n^2}{2} 
\end{equation}
and leads to
\begin{equation}
\label{monomial_solution}
    \left(\frac{\phi}{M_p}\right)^2 \sim 2nN + \frac{n^2}{2}.
\end{equation}
The fundamental result in Equation~\eqref{monomial_solution} provides the expression 
\begin{equation}
    P_S \sim \frac{1}{12\pi^2 n^2}\frac{M^4_{\text{inf}}}{M^4_p}
    \left(2nN + n^2/2\right)^{1 + n/2}
\end{equation}
for the scalar spectrum. It can be inverted in terms of the ratio $M_{\text{inf}}/M_p$ and, if one applies the current estimate for the scalar spectrum amplitude of the COBE mission, $P_S^{\text{cobe}}\sim 2\times 10^{-9}$, it is possible to 
constrain the inflationary scale for a given $n$ as
\begin{equation}
    \frac{M_{\text{inf}}}{M_p} \sim \left[\frac{12\pi^2 n^2 P_S^{\text{cobe}}}{\left(2 n N + n^2/2\right)^{1+n/2}}\right]^{1/4}.
\end{equation}
In principle, one can assume that inflation occurs at some specific scale, for example, around \mbox{$10^{15}$ GeV}.  
This statement, together with the COBE outcome,  offers the possibility to constrain the model parameter $n$.
At the same time, the expressions for the set of inflationary parameters in terms of $N$ are
\begin{align}
    n_S(N)&\sim 1 - \frac{n(n+2)}{2nN + n^2/2},\\
    r(N) &\sim \frac{8n^2}{2nN + n^2/2},\\
    \alpha_S &\sim -\frac{2n^2(n+2)}{(2nN + n^2/2)^2},\\ 
    \alpha_T &\sim \frac{2n^3}{(2nN + n^2/2)^2}.
\end{align}

To give an example, selecting the massive case ($n=2$) that one has at first order leads to
\begin{align}
    n_S(N) &\sim 1 - \frac{2}{N}\\
    r(N)&\sim \frac{8}{N}\\
    \alpha_S &\sim -\frac{2}{N^2}\\
    \alpha_T &\sim \frac{1}{N^2}.
\end{align}
If one assumes that the exits of reference scales correspond to $N\sim 60$, then 
the related inflaton value is of the order of $\phi\sim 15 M_p$, 
the predicted reference energy scale is $M_{\text{inf}}\sim 2\times 10^{15}$ GeV,
and, as a consequence, the inflaton mass is estimated to be \mbox{$m_{\phi}\sim 10^{12}$ GeV.} 
{The numerical predictions for the two main inflationary observables result in}
\begin{eqnarray}
    N\sim 60, \quad n_S \sim 0.9667 \quad r\sim 0.1333  \\
    N\sim 50, \quad n_S \sim 0.9600 \quad r\sim 0.1600
\end{eqnarray}

It is worth to stress that the numerical solution of the complete Equation~\eqref{inflaton_equation_efolds}
provides slightly different (and clearly more precise) results.
{For example, for $N\sim 60$, one obtains $n_S\sim 0.9669$ and $r\sim 0.1322$,
while for $N\sim 50$, $n_S\sim 0.9606$ and $r\sim 0.1573$}.

The same recipe can be used to evaluate other scenarios, as, for example, the model with $n=4$.
In the single case of $N\sim 60$, the scalar field value of the quartic case is $\phi\sim 20 M_p$ and the related energy scale is 
$M_{\text{inf}} \sim 10^{14}$ GeV, with a very small self-coupling constant $\lambda_{\phi}\sim 10^{-15}-10^{-16}$.
Unfortunately, both the quadratic and quartic cases are not compatible with the current Planck and BICEP results.

The monomial scenario provides a very simple function $N(\phi)$ that can be easily inverted to give the solution $\phi(N)$. 
Moreover, a similar discussion can be performed for the exponential scenario (see~\cite{Lucchin:1984yf,Halliwell:1986ja,Burd:1988ss}),
where an exact solution for $\phi(N)$ and $a(t)$ can also be obtained.
Unfortunately, this is not always the case. For instance, there are models in which the inversion must proceed via some special functions, preventing a simple analytical interpretation for the solution $\phi(N)$.
An interesting example is given by the class of the so-called $\alpha$-attractor models. They can be
generated in several different ways, although the most advanced version emerges from supergravity  
\cite{Kallosh:2013hoa,Kallosh:2013daa,Ferrara:2013rsa,Kallosh:2013xya,Kallosh:2013yoa,Galante:2014ifa,Kallosh:2015zsa,Kallosh:2016gqp}).  
The $\alpha$-attractor potentials are typically divided into two subclasses
called E models and T models, both characterized by a flat region (for $\phi>M_p$) that comes out to be protected against quantum corrections~\cite{Kallosh:2016gqp}.
In the present review, the focus will be limited only to the E-models, whose scalar potential is
\begin{equation}
    V(\phi) = M^4_{\text{inf}} \left( 1 - e^{-b\phi/M_p} \right)^{2n}, \quad b = \sqrt{\frac{2}{3\alpha}},
\end{equation}
with the global minimum located at $\phi=0$.
The scalar potential is characterized by a couple of free parameters, $n$ and $\alpha$.
In supergravity, the $\alpha$ parameter is related to the K\"ahler curvature of the moduli space manifold, where the inflaton superfield is a coordinate~\cite{Kallosh:2015zsa}.
It is important to stress how this family of potentials naturally reproduces well-known inflationary models.
For example, the case $n=1$ corresponds to the Einstein frame version of the Starobinsky model of inflation~\cite{Maeda:1987xf}, while the case
$\alpha=1/9$ is the supergravity Linde--Goncharov model~\cite{Goncharov:1983mw,Goncharov:1984jlb}.
It should also be noticed that for large values of $b$ (or $\alpha\ll1$), the exponential term tends to be suppressed and the scalar function
approximates a step-behavior.
On the contrary, for small values of $b$ (or $\alpha\gg1$), $V(\phi)$ tends to assume a simple quadratic parabolic shape.
In order to simplify the treatment, one can simply consider the scenario with $n=1$.
In this case, the derivatives of the potential
\begin{equation}
 V^{(k)}(\phi) = \frac{M^4_{\text{inf}}}{M_p^k} (-1)^{k-1} 2b^k e^{-b\phi/M_p}\left(1 - 2^{k-1}e^{-b\phi/M_p}\right) 
\end{equation}
allow us to write down the PRSPs.
In particular, the first two comes out to be
\begin{eqnarray}
    \epsilon_V(\phi) &=& \frac{2b^2}{\left(e^{b\phi/M_p} - 1\right)^2} , \\
    \eta_V(\phi) &=& -2b^2\frac{\left(e^{b\phi/M_p} - 2\right)}{\left(e^{b\phi/M_p} - 1\right)^2} .
\end{eqnarray}
As a consequence, one obtains
\begin{equation}
    P_S(\phi) \sim \frac{1}{48\pi^2}\left(\frac{M_{\text{inf}}}{M_p}\right)^4
    \frac{e^{2b\phi/M_p}}{b^2}\left(1 - e^{-b\phi/M_p}\right)^4
\end{equation}
and 
\begin{eqnarray}
    n_S(\phi) &\sim& 1 - 4b^2\frac{\left(e^{b\phi/M_p} + 1\right)}{\left(e^{b\phi/M_p} - 1\right)^2} ,\\
    r(\phi) &\sim& \frac{32b^2}{\left(e^{b\phi/M_p} - 1\right)^2} .
\end{eqnarray}
The solution of Equation~\eqref{inflaton_efolds_approx} gives
\begin{equation}
    N(\phi)\sim \frac{1}{2b^2}\left( e^{b\phi/M_p} - e^{b\phi_{end}/M_p}  \right) - 
    \frac{1}{2b}\left( \frac{\phi}{M_p} - \frac{\phi_{end}}{M_p}  \right) ,
\label{EnneInAlfaAtt}\end{equation}
where $\phi_{\text{end}}$, defined by  $\epsilon_V(\phi_{end})\sim 1$,  results in
\begin{equation}
    \frac{\phi_{end}}{M_p}\sim \frac{1}{b}\log(1 + \sqrt{2}b) .
\end{equation}
The inversion of the expression in Equation~\eqref{EnneInAlfaAtt} is manifestly not a simple task.
One can try to use to the Lambert function $W$~\cite{Martin:2013tda} to obtain
\begin{equation}
\frac{\phi(N)}{M_p} = \frac{1}{b}\left[-\mathcal{F}(N) - W(-e^{-\mathcal{F}(N)})\right]    , 
\end{equation}
where  
\begin{equation}
    \mathcal{F}(N) = 2b^2N + e^{b\phi_{end}/M_p} - b\frac{\phi_{\text{end}}}{M_p} .
\end{equation}
Unfortunately, this result does not provide a straightforward interpretation of the field evolution.
It is thus convenient to adopt an additional approximation, assuming that the plateau inflaton value is larger than the final field value,  $\phi\gg\phi_{\text{end}}$.
It implies that
\begin{equation}
    e^{b\phi/M_p}\sim 2b^2 N, \quad \phi(N)\sim \frac{1}{b}\ln{2b^2N} ,
\end{equation}
giving rise to the scalar power spectrum
\begin{equation}
    P_S \sim \frac{N^2}{18\alpha\pi^2}\left(\frac{M_{\text{inf}}}{M_p}\right)^4 ,
\end{equation}
whose inversion for the inflation scale provides
\begin{equation}
    \frac{M_{\text{inf}}}{M_p} \sim \frac{18\alpha\pi^2}{N^2} P_S^{cobe} . 
\end{equation}

On the other hand, the inflationary observables decrease with $N$ as
\begin{eqnarray}
    n_S(N) &\sim& 1 - 4b^2e^{-b\phi(N)/M_p} + ... = 1 - \frac{2}{N}, \\
    r(N) &\sim& 32b^2e^{-2b\phi(N)/M_p} + ... = \frac{12\alpha}{N^2} .
\end{eqnarray}
The simplest example to discuss is the Starobinsky model corresponding, as mentioned, to $n=1$.
Taking the standard horizon crossing option $N\sim 60$, one has a scalar field at horizon crossing $\phi\gtrsim 5$$M_p$,
an inflationary reference scale $M_{\text{inf}}\sim 3\times 10^{15}$ GeV, and numerical estimates for scalar tilt and tensor-to-scalar ratio:
\begin{eqnarray}
    N\sim 60, \quad n_S \sim 0.9667, \quad r\sim 0.0033 \\
    N\sim 50, \quad n_S \sim 0.9600, \quad r\sim 0.0048.
\end{eqnarray}
The numerical investigation would reveal $n_S\sim 0.9653$ and $r\sim 0.0034$ for $N\sim 60$ and $n_S\sim 0.9584$ with $r\sim 0.0049$ for $N\sim 50$.
Therefore, the Starobinsky case as well as all the $\alpha$-attractor cases with a small model parameter (i.e., $\alpha<1$) are extremely compatible with the current observational constraints.

To conclude this very brief overview, it can be interesting to analyze models  where the integral giving $N(\phi)$ can be exactly performed, but the resulting expression cannot be explicitly inverted, even resorting to special functions.
In this context, a recently discovered class of models, based on an Einstein--Cartan extension of general relativity, warrants analysis~\cite{Pradisi:2022nmh, Salvio:2022suk, DiMarco:2023ncs}. 
In particular, it has been shown that one may consider a gravity theory whose action is a combination of the scalar curvature term and a non-linear function of the so-called parity violating Holst invariant term:
\begin{equation}
    {\cal R'} \equiv \varepsilon^{\mu\nu\rho\sigma}{\cal R}_{\mu\nu\rho\sigma},
\end{equation}
where ${\cal R}_{\mu\nu\rho\sigma}$ is the curvature tensor and $\varepsilon^{\mu\nu\rho\sigma}$ is the Levi--Civita invariant tensor. $ {\cal R'}$ obviously vanishes in a Riemannian geometry, but in a Palatini approach, it provides a dynamical torsion and results in classically equivalent (on shell) to a pseudoscalar field, minimally coupled to general relativity, and equipped with a non-trivial potential suitable to drive a single-field slow-roll inflationary phase.  Indeed, if $\phi$ indicates the pseudoscalar field unequivocally emerging from the underlying non-Riemannian geometry, the induced potential comes out to be
\begin{equation}
    V(\phi) = M^4_{\text{inf}} \ \Bigg|\gamma\sinh X(\phi)-1\Bigg|^\frac{p}{p-1} , 
\label{eq:pseudopotential}\end{equation}
with
\begin{equation}
    X(\phi) = \sqrt{\frac{2}{3}} \frac{\phi}{M_p} + \theta_{\gamma}
\end{equation}
and the reference scale of inflation is given by
\begin{equation}
    M^4_{\text{inf}} = \frac{p-1}{p^{p/(p-1)}}\frac{1}{\xi^{\frac{1}{p-1}}}\Bigg|\frac{M^2_p}{4\gamma}\Bigg|^\frac{p}{p-1} .
\end{equation}
In the previous expressions (for details, see~\cite{DiMarco:2023ncs} and refeences therein), $\gamma$ is the so-called Barbero--Immirzi parameter, $p \geq 1$ is a real parameter, $\xi$ is the coupling of the non-linear term to the metric, and 
\begin{equation}
    \theta_{\gamma} = \sinh^{-1}(\gamma^{-1}) .
\end{equation}
The sign of the Barbero--Immirzi parameter determines the direction of the slow-roll phase.
Specifically, the slow-roll phase occurs for decreasing values of the inflaton field (i.e., $\dot{\phi}<0$)
for negative values of $\gamma$, while it occurs for increasing values of $\phi$
(i.e., $\dot{\phi}>0$) for positive values of $\gamma$.
The first derivative of the scalar potential reads
\begin{equation}
    V'(\phi) = \frac{M^4_{\text{inf}}}{M_p} \ \dfrac{\sqrt{2} \, \gamma \,  p\, \cosh{X(\phi)}\left|\gamma\sinh{X(\phi)}-1\right|^\frac{p}{p-1}}{\sqrt{3}\left(p-1\right)\left(\gamma\sinh{X(\phi)}-1\right)} 
\end{equation}
and the second derivative is
\begin{equation}
    V''(\phi) = \frac{M^4_{\text{inf}}}{M^2_p} \ \frac{2 \, \gamma \, p \, \left|\gamma\sinh{X(\phi)}-1\right|^\frac{p}{p-1}}{3 (p-1)\left(\gamma\sinh{X(\phi)}-1\right)^2}
    \left(\frac{\gamma}{p-1}(p\sinh^2{X(\phi)}+1)-\sinh{X(\phi)}\right) .
\end{equation}
Thus, the firsts two PSRPs can be calculated and result in
\begin{equation}
    \epsilon_V(\phi) = 
    \frac{\gamma^2}{3}\left(\frac{p}{p-1}\right)^2
    \frac{\cosh^2{X(\phi)}}{\left(\gamma\sinh{X(\phi)}-1\right)^2}
\label{epsilonV}\end{equation}
and
\begin{equation}
    \eta_V(\phi) = 
    \frac{2\, \gamma \, p}{3 (p-1)\left(\gamma\sinh{X(\phi)}-1\right)^2}
    \left(\frac{\gamma}{p-1}(p\sinh^2{X(\phi)}+1)-\sinh{X(\phi)}\right) .
\end{equation}
The power spectrum assumes the very complicated expression
\begin{equation}
    P_S(\phi)\sim \frac{M^4_{\text{inf}}}{18 \, \pi^2 \, M^4_p \, \gamma^2} \left(\frac{p-1}{p}\right)^2
    \frac{\left(\gamma\sinh{X(\phi)}-1\right)^2}{\cosh^2{X(\phi)}}
    \Bigg|\gamma\sinh{X(\phi)}-1\Bigg|^\frac{p}{p-1} ,
\label{scalarpower}\end{equation}
while the two main observables turn out to be
\begin{eqnarray}
    n_S(\phi) &\sim& 
    1 - \frac{2 \, \gamma^2}{3}\left(\frac{p}{p-1}\right)^2\frac{\cosh^2{X(\phi)}}{\left(\gamma\sinh{X(\phi)}-1\right)^2} \nonumber\\
    &+& \frac{4 \, \gamma \, p}{3 (p-1)\left(\gamma\sinh{X(\phi)}-1\right)^2}\left[\frac{\gamma}{p-1}\left(p\sinh^2{X(\phi}, - 1\right)-\sinh{X(\phi)}\right]
\end{eqnarray}
and
\begin{equation}
    r(\phi) \sim \frac{16}{3} \, \gamma^2\left(\frac{p}{p-1}\right)^2 \frac{\cosh^2{X(\phi)}}{\left(\gamma\sinh{X(\phi)}-1\right)^2} .
\end{equation}
As usual, to obtain first-order expressions of the observables and an estimate of the model parameter $M_{\text{inf}}$, the solution of the integral $N(\phi)$ is required. 
One obtains
\begin{align}
    \label{slow-roll solution}
    &N(\phi) =
    \frac{3(p-1)}{2 p}
    \left[
    \ln{\Bigg|\cosh{X(\phi)}\Bigg|}
    - \frac{1}{\gamma}\tan^{-1}\left(\sinh{X(\phi)}\right)
    \right]\Bigg|_{\phi_{end}}^\phi \nonumber\\
    &=
     \frac{3(p-1)}{2 p}
    \ln{\frac{\Bigg|\cosh{X(\phi)}\Bigg|}{\Bigg| \cosh{X(\phi_{end})} \Bigg|}}
    -  \frac{3(p-1)}{2 \, \gamma \, p}
    \left[\tan^{-1}\left(\sinh{X(\phi)}\right) - 
    \tan^{-1}\left(\sinh{X(\phi_{end})}\right)\right] ,
\end{align}
which neatly show how there is no hope to analytically determine $\phi(N)$ by inversion, unless a very specific range of the Barbero--Immirzi parameter ($|\gamma| \gg 10^{-1}$) is considered. However, the function in Equation~\eqref{slow-roll solution} is still very useful to extract possible pairs $(N,\phi)$, allowing us to derive simple numerical predictions for the cosmological observables $\mathcal{O}(\phi)$.
For instance, in the case $p=2$, choosing $\gamma\sim 10^{-3}$ and $N\sim 60$, one obtains
\begin{equation}
    n_S \sim 0.9680, \quad r\sim 0.003 ,
\end{equation}
with a very high inflation reference scale $M_{\text{inf}} \sim 5.6\times 10^{15}$ GeV.
Such predictions result in a very good agreement with the current constraints.

A final observation is in order: there are cases, of course, where the potential function $V(\phi)$ does not allow for an exact calculation of the integral $N(\phi)$.
In a similar situation, it is convenient to proceed by approximating the potential with its slow-roll expression.
It happens, for instance, in superstring-inspired models  like the ones called fiber-inflation scenarios 
\cite{Cicoli:2008va,Cicoli:2008gp,Cicoli:2016xae,Cicoli:2020bao}.

\section{Summary and Conclusions}
\label{summary and conclusions}

The conventional Einstein--Friedmann--Klein--Gordon system of equations offers the simplest framework for exploring the evolution of the single-field slow-roll inflationary scenario in the early universe. 
Moreover, the solutions for the scale factor $a(t)$ and the field trajectory $\phi(t)$ have a straightforward interpretation
and provide insights into the overall dynamics. 
However, the time-dependent profile of the inflaton solution proves to be less practical for tasks like the computation of inflationary observables that are, quite naturally, functions of the scalar field $\phi$ via potential slow-roll parameters (PRSPs) and typically exhibit an unusable dependence upon time. In light of this, it becomes unavoidable to reformulate the inflaton equation using an alternative and more suitable time-like variable. The number of $e$-folds before the end of inflation, denoted by $N$ and tracking the times the universe grows exponentially until the end of the slow-roll phase, appears to be the most natural.  
Rough evaluations suggest that its value from the horizon crossing (of the observable perturbed scales)
to the end of inflation typically falls within the range between $50$ and $60$.
Actually, many more possibilities exist, also depending on the properties of the postinflationary phase (see, for example,~\cite{Das:2015uwa,Cicoli:2016olq,Maharana:2017fui,DiMarco:2018bnw}).

In this paper, we have reviewed the derivation and uses of 
the inflaton equations based on $N$ as a fundamental time-like variable in an approach that is widely used and appreciated in the current literature, especially in order to obtain deeper insights on the inflationary dynamics. 
The procedure is characterized by the introduction of a new form of the first slow-roll parameter $\epsilon$, which is
slightly different from the traditional one.

Complete numerical solutions of the new equation require initial conditions that are suited to describe the horizon crossing of the relevant cosmologically perturbed scales. Alternatively, one can apply the slow-roll approximation, which returns the standard definition of $N$ and facilitates the computation of a first-order, analytical field solution $\phi(N)$, which is extremely useful to rephrase the dependency {of the standard inflationary observables} in terms of $N$.

Finally, it is worth emphasizing another point that underscores the importance of the number of $e$-folds. 
In Section~\ref{slow-roll parameters, inflationary perturbations and observables}, the focus was on the theory of linear or first-order metric perturbations. 
However, as is well known, it is also important to study the behavior of non-linear curvature perturbations 
since they can provide additional inflation observables like higher-order spectra
(bispectrum, trispectrum$\ldots$), which are useful to estimate non-Gaussianities on CMB scales (see, for instance, 
\cite{Acquaviva:2002ud,Maldacena:2002vr,Bartolo:2004if,Babich:2004gb,Chen:2010xka,Arkani-Hamed:2015bza}).
In those situations, the standard computation approach proves to be quite cumbersome. 
More suitable non-perturbative methods have thus been introduced, such as the $\delta N$ formalism 
\cite{Sasaki:1995aw,Lyth:2004gb,Wands:2000dp,Dias:2012qy,Sugiyama:2012tj}, based on counting the local number of $e$-folds of various local patches. The $\delta N$ formalism is widely used in the current literature due to its easy-to-use nature (for a recent review, see~\cite{Abolhasani:2019cqw}).

\acknowledgments
We would like to thank G. De Gasperis and Y. Mambrini for interesting discussions. E.O. would like to thank the Physics Department of the University of Rome ``Tor Vergata'' for the kind hospitality while this work was being undertaken. E.O. was partially supported by the ``Bando Visiting Professor 2023,  University of Rome ``Tor Vergata'', Prot. n. 0038224 (26/07/2023), by the project PROMETEO/2020/079 (Generalitat Valenciana),  CNPq (Brazil), grant No.314392/2021-1 and by the Spanish Agencia Estatal de Investigacion Grant No. PID2022-138607NB-I00, funded by 14 MCIN/AEI/10.13039/501100011033, FEDER, UE, and ERDF {("A way of making Europe").}

\appendix

\bibliographystyle{apsrev4-2}

\begin{thebibliography}{999}


\bibitem{Starobinsky:1980te}
Starobinsky, A.A.
A New Type of Isotropic Cosmological Models Without Singularity. \emph{Phys. Lett. B} \textbf{1980}, \emph{91}, 99--102.
\url{http://doi.org/10.1016/0370-2693(80)90670-X}.


\bibitem{Guth:1980zm}
Guth, A.H.
The Inflationary Universe: A Possible Solution to the Horizon and Flatness Problems. \emph{Phys. Rev. D} {\bf 1981}, \emph{23}, 347.
\url{http://doi.org/10.1103/PhysRevD.23.347}.

\bibitem{Linde:1981mu}
Linde, A.D.
A New Inflationary Universe Scenario: A Possible Solution of the Horizon, Flatness, Homogeneity, Isotropy and Primordial Monopole Problems. \emph{Phys.\ Lett.}  {\bf 1982}, \emph{108B}, 389.
\url{http://doi.org/10.1016/0370-2693(82)91219-9}.

\bibitem{Albrecht:1982wi}
Albrecht, A.; Steinhardt, P.J.
Cosmology for Grand Unified Theories with Radiatively Induced Symmetry Breaking. \emph{Phys.\ Rev.\ Lett.\ } {\bf 1982}, \emph{48}, 1220.
\url{http://doi.org/10.1103/PhysRevLett.48.1220}.

\bibitem{Hawking:1981fz}
Hawking, S.W.; Moss, I.G.
Supercooled Phase Transitions in the Very Early Universe.
\emph{Phys.\ Lett.\  }{\bf 1982}, \emph{110B}, 35.
\url{http://doi.org/doi:10.1016/0370-2693(82)90946-7}.

\bibitem{Linde:1983gd}
Linde, A.D.
Chaotic Inflation.
\emph{Phys.\ Lett.\  }{\bf 1983}, \emph{129B}, 177.
\url{http://doi.org/10.1016/0370-2693(83)90837-7}.


\bibitem{Linde:1990flp}
Linde, A.D.
Particle physics and inflationary cosmology. \emph{Contemp. Concepts Phys.} \textbf{1990}, \emph{5}, 1--362.
\url{http://arxiv.org/abs/hep-th/0503203}.

\bibitem{Linde:2007fr}
Linde, A.D.
Inflationary Cosmology.
\emph{Lect. Notes Phys.} \textbf{2008}, \emph{738}, 1--54.
\url{http://doi.org/10.1007/978-3-540-74353-8\_1}.

\bibitem{Olive:1989nu} 
Olive, K.A.
Inflation.
\emph{Phys.\ Rept.\ } {\bf 1990}, \emph{190}, 307.
\url{http://doi.org/10.1016/0370-1573(90)90144-Q}.

\bibitem{Baumann:2009ds}
Baumann, D.
INFLATION.  \emph{Physics of the Large and the Small} \textbf{2011}, 
523-686.
\url{http://doi.org/10.1142/9789814327183\_0010}





\bibitem{Kolb:1990vq}
Kolb, E.W.; Turner, M.S.
The Early Universe.
\emph{Front. Phys.} \textbf{1990}, \emph{69}, 1--547
\url{http://doi.org/10.1201/9780429492860}.

\bibitem{Mukhanov:2005sc}
Mukhanov, V.
Physical Foundations of Cosmology.
Cambridge University Press: Cambridge, UK, 
 2005; ISBN 978-0-521-56398-7.
\url{http://doi.org/10.1017/CBO9780511790553}.

\bibitem{Weinberg:2008zzc} 
Weinberg, S.
 \emph{ Cosmology};
Oxford University Press: Oxford, UK, 2008; 593p.


\bibitem{Gorbunov:2011zzc} 
Gorbunov, D.S.; Rubakov, V.A.
\emph{Introduction to the Theory of the Early Universe: Cosmological Perturbations and Inflationary Theory};
World Scientific: Hackensack, NJ, USA, 2011; 489p.
\url{http://doi.org/10.1142/7874}.

\bibitem{Baumann:2018muz}
Baumann, D.
Primordial Cosmology.
\emph{PoS} \textbf{2018}, \emph{TASI2017}, 009.
\url{http://doi.org/10.22323/1.305.0009}


\bibitem{Mukhanov:1981xt}
Mukhanov, V.F.; Chibisov, G.V.
Quantum Fluctuations and a Nonsingular Universe.
\emph{JETP Lett.} \textbf{1981}, \emph{33}, 532--535.

\bibitem{Hawking:1982cz}
Hawking, S.W.
The Development of Irregularities in a Single Bubble Inflationary Universe.
\emph{Phys. Lett. B} \textbf{1982}, \emph{115}, 295.
\url{http://doi.org/10.1016/0370-2693(82)90373-2}.


\bibitem{Starobinsky:1982ee}
Starobinsky, A.A.
Dynamics of Phase Transition in the New Inflationary Universe Scenario and Generation of Perturbations.
\emph{Phys. Lett. B} \textbf{1982}, \emph{117}, 175--178.
\url{http://doi.org/10.1016/0370-2693(82)90541-X}.

\bibitem{Guth:1982ec}
Guth, A.H.; Pi, S.Y.
Fluctuations in the New Inflationary Universe.
\emph{Phys. Rev. Lett. }\textbf{1982}, \emph{49}, 1110--1113.
\url{http://doi.org/10.1103/PhysRevLett.49.1110}.


\bibitem{Linde:1982uu}
Linde, A.D.
Scalar Field Fluctuations in Expanding Universe and the New Inflationary Universe Scenario.
\emph{Phys. Lett. B} \textbf{1982}, \emph{116}, 335--339.
\url{http://doi.org/10.1016/0370-2693(82)90293-3}.


\bibitem{Bardeen:1983qw}
Bardeen, J.M.; Steinhardt, P.J.; Turner, M.S.
Spontaneous Creation of Almost Scale---Free Density Perturbations in an Inflationary Universe.
\emph{Phys. Rev. D} \textbf{1983}, \emph{28}, 679.
\url{http://doi.org/10.1103/PhysRevD.28.679}.

\bibitem{Lyth:1984gv}
Lyth, D.H.
Large Scale Energy Density Perturbations and Inflation.
\emph{Phys. Rev. D} \textbf{1985}, \emph{31}, 1792--1798.
\url{http://doi.org/10.1103/PhysRevD.31.1792}.


\bibitem{Mukhanov:1985rz}
Mukhanov, V.F.
Gravitational Instability of the Universe Filled with a Scalar Field.
\emph{JETP Lett.} \textbf{1985}, \emph{41}, 493--496.


\bibitem{Sasaki:1986hm}
Sasaki, M.
Large Scale Quantum Fluctuations in the Inflationary Universe.
\emph{Prog. Theor. Phys. }\textbf{1986}, \emph{76}, 1036.
\url{http://doi.org/10.1143/PTP.76.1036}.

\bibitem{Stewart:1993bc}
Stewart, E.D.; Lyth, D.H.
A More accurate analytic calculation of the spectrum of cosmological perturbations produced during inflation.
\emph{Phys. Lett. B} \textbf{1993}, \emph{302}, 171--175.
\url{http://doi.org/10.1016/0370-2693(93)90379-V}.


\bibitem{Grishchuk:1974ny}
Grishchuk, L.P.
Amplification of gravitational waves in an istropic universe.
\emph{Zh. Eksp. Teor. Fiz.} \textbf{1974}, \emph{67}, 825--838.


\bibitem{Starobinsky:1979ty}
Starobinsky, A.A.
Spectrum of relict gravitational radiation and the early state of the universe.
\emph{JETP Lett.} \textbf{1979}, \emph{30}, 682--685.


\bibitem{Rubakov:1982df}
Rubakov, V.A.; Sazhin, M.V.; Veryaskin, A.V.
Graviton Creation in the Inflationary Universe and the Grand Unification Scale.
\emph{Phys. Lett. B} \textbf{1982}, \emph{115}, 189--192.
\url{http://doi.org/10.1016/0370-2693(82)90641-4}.

\bibitem{Fabbri:1983us}
Fabbri, R.; Pollock, M.d.
The Effect of Primordially Produced Gravitons upon the Anisotropy of the Cosmological Microwave Background Radiation.
\emph{Phys. Lett. B} \textbf{1983}, \emph{125}, 445--448.
\url{http://doi.org/10.1016/0370-2693(83)91322-9}.

\bibitem{Abbott:1984fp}
Abbott, L.F.; Wise, M.B.
Constraints on Generalized Inflationary Cosmologies.
\emph{Nucl. Phys. B} \textbf{1984}, \emph{244}, 541--548.
\url{http://doi.org/10.1016/0550-3213(84)90329-8}.

\bibitem{Allen:1987bk}
Allen, B.
The Stochastic Gravity Wave Background in Inflationary Universe Models.
\emph{Phys. Rev. D} \textbf{1988}, \emph{37}, 2078.
\url{http://doi.org/10.1103/PhysRevD.37.2078}.

\bibitem{Lucchin:1992qi}
Lucchin, F.; Matarrese, S.; Mollerach, S.
The Gravitational wave contribution to CMB anisotropies and the amplitude of mass fluctuations from COBE results.
\emph{Astrophys. J. Lett.} \textbf{1992}, \emph{401}, L49.
\url{http://doi.org/10.1086/186668}.

\bibitem{Turner:1993vb}
Turner, M.S.; White, M.J.; Lidsey, J.E.
Tensor perturbations in inflationary models as a probe of cosmology.
\emph{Phys. Rev. D} \textbf{1993}, \emph{48}, 4613--4622.
\url{http://doi.org/10.1103/PhysRevD.48.4613}.

\bibitem{Crittenden:1993wm}
Crittenden, R.; Davis, R.L.; Steinhardt, P.J.
Polarization of the microwave background due to primordial gravitational waves.
\emph{Astrophys. J. Lett.} \textbf{1993}, \emph{417}, L13--L16.
\url{http://doi.org/10.1086/187082}.

\bibitem{Knox:1994qj}
Knox, L.; Turner, M.S.
Detectability of tensor perturbations through CBR anisotropy.
\emph{Phys. Rev. Lett. }\textbf{1994}, \emph{73}, 3347--3350.
\url{http://doi.org/10.1103/PhysRevLett.73.3347}.

\bibitem{Turner:1996ck}
Turner, M.S.
Detectability of inflation produced gravitational waves.
\emph{Phys. Rev. D} \textbf{1997}, \emph{55}, R435--R439.
\url{http://doi.org/10.1103/PhysRevD.55.R435}.



\bibitem{Mukhanov:1990me}
Mukhanov, V.F.; Feldman, H.A.; Brandenberger, R.H.
Theory of cosmological perturbations. Part 1. Classical perturbations. Part 2. Quantum theory of perturbations. Part 3. Extensions.
\emph{Phys. Rept.} \textbf{1992}, \emph{215}, 203--333.
\url{http://doi.org/10.1016/0370-1573(92)90044-Z}.

\bibitem{Riotto:2002yw}

Riotto, A.
Inflation and the theory of cosmological perturbations.
\emph{ICTP Lect. Notes Ser.} \textbf{2003}, \emph{14}, 317--413.
\url{http://arxiv.org/abs/hep-ph/0210162}{[hep-ph]}.

\bibitem{Guzzetti:2016mkm}
Guzzetti, M.C.; Bartolo, N.; Liguori, M.; Matarrese, S.
Gravitational waves from inflation.
\emph{Riv. Nuovo Cim.} \textbf{2016}, \emph{39}, 399--495.
\url{http://doi.org/10.1393/ncr/i2016-10127-1}.



\bibitem{Steinhardt:1984jj}
Steinhardt, P.J.; Turner, M.S.
A Prescription for Successful New Inflation.
\emph{Phys. Rev. D} \textbf{1984}, \emph{29}, 2162--2171
\url{http://doi.org/10.1103/PhysRevD.29.2162}.

\bibitem{Liddle:1992wi}
Liddle, A.R.; Lyth, D.H.
COBE, gravitational waves, inflation and extended inflation.
\emph{Phys. Lett. B} \textbf{1992}, \emph{291}, 391--398.
\url{http://doi.org/10.1016/0370-2693(92)91393-N}.

\bibitem{Liddle:1994dx}
Liddle, A.R.; Parsons, P.; Barrow, J.D.
Formalizing the slow roll approximation in inflation.
\emph{Phys. Rev. D} \textbf{1994}, \emph{50}, 7222--7232.
\url{http://doi.org/10.1103/PhysRevD.50.7222}.


\bibitem{Martin:2013tda}
Martin, J.; Ringeval, C.; Vennin, V.
Encyclop\ae{}dia Inflationaris.
\emph{Phys. Dark Univ.} \textbf{2014}, \emph{5--6}, 75--235.
\url{http://doi.org/10.1016/j.dark.2014.01.003}.



\bibitem{Coughlan:1985mw}
Coughlan, G.D.; Ross, G.G.
Initial Conditions for Inflation.
\emph{Phys. Lett. B} \textbf{1985}, \emph{157}, 151--156.
\url{http://doi.org/10.1016/0370-2693(85)91536-9}.

\bibitem{Linde:1985ub}
Linde, A.D.
Initial conditions for inflation.
\emph{Phys. Lett. B} \textbf{1985}, \emph{162}, 281--286.
\url{http://doi.org/10.1016/0370-2693(85)90923-2}.

\bibitem{Albrecht:1984qt}
Albrecht, A.; Brandenberger, R.H.
On the Realization of New Inflation.
\emph{Phys. Rev. D} \textbf{1985}, \emph{31}, 1225.
\url{http://doi.org/10.1103/PhysRevD.31.1225}.


\bibitem{Albrecht:1985yf}
Albrecht, A.; Brandenberger, R.H.; Matzner, R.
Numerical Analysis of Inflation.
\emph{Phys. Rev. D} \textbf{1985}, \emph{32}, 1280.
\url{http://doi.org/doi:10.1103/PhysRevD.32.1280}.

\bibitem{Albrecht:1986pi}
Albrecht, A.; Brandenberger, R.H.; Matzner, R.
Inflation with Generalized Initial Conditions.
\emph{Phys. Rev. D} \textbf{1987}, \emph{35}, 429.
\url{http://doi.org/10.1103/PhysRevD.35.429}.

\bibitem{Kurki-Suonio:1987mrt}
Kurki-Suonio, H.; Matzner, R.A.; Centrella, J.; Wilson, J.R.
Inflation From Inhomogeneous Initial Data in a One-dimensional Back Reacting Cosmology.
\emph{Phys. Rev. D} \textbf{1987}, \emph{35}, 435--448.
\url{http://doi.org/10.1103/PhysRevD.35.435}.
\bibitem{Goldwirth:1989pr}
Goldwirth, D.S.; Piran, T.
Inhomogeneity and the Onset of Inflation.
\emph{Phys. Rev. Lett. }\textbf{1990}, \emph{64}, 2852--2855.
\url{http://doi.org/10.1103/PhysRevLett.64.2852}.

\bibitem{Goldwirth:1991rj}
Goldwirth, D.S.; Piran, T.
Initial conditions for inflation.
\emph{Phys. Rept.} \textbf{1992}, \emph{214}, 223--291.
\url{http://doi.org/10.1016/0370-1573(92)90073-9}.

\bibitem{Kurki-Suonio:1993lzy}
Kurki-Suonio, H.; Laguna, P.; Matzner, R.A.
Inhomogeneous inflation: Numerical evolution.
\emph{Phys. Rev. D} \textbf{1993}, \emph{48}, 3611--3624.
\url{http://doi.org/10.1103/PhysRevD.48.3611}.


\bibitem{Iguchi:1996rh}
Iguchi, O.; Ishihara, H.
Onset of inflation in inhomogeneous cosmology.
\emph{Phys. Rev. D} \textbf{1997}, \emph{56}, 3216--3224.
\url{http://doi.org/10.1103/PhysRevD.56.3216}.

\bibitem{Vachaspati:1998dy}
Vachaspati, T.; Trodden, M.
Causality and cosmic inflation.
\emph{Phys. Rev. D} \textbf{1999}, \emph{61}, 023502.
\url{http://doi.org/10.1103/PhysRevD.61.023502}.

\bibitem{Easther:2014zga}
Easther, R.; Price, L.C.; Rasero, J.
Inflating an Inhomogeneous Universe.
\emph{JCAP} \textbf{2014}, \emph{8}, 41.
\url{http://doi.org/10.1088/1475-7516/2014/08/041}.

\bibitem{East:2015ggf}
East, W.E.; Kleban, M.; Linde, A.; Senatore, L.
Beginning inflation in an inhomogeneous universe.
\emph{JCAP} \textbf{2016}, \emph{9}, 10.
\url{http://doi.org/10.1088/1475-7516/2016/09/010}.




\bibitem{Lyth:1996im} 
Lyth, D.H.
What would we learn by detecting a gravitational wave signal in the cosmic microwave background anisotropy?
\emph{Phys.\ Rev.\ Lett.\ } {\bf 1997}, \emph{78}, 1861.
\url{http://doi.org/10.1103/PhysRevLett.78.1861}.

\bibitem{Efstathiou:2005tq} 
Efstathiou, G.; Mack, K.J.
The Lyth bound revisited.
\emph{JCAP} {\bf 2005}, \emph{008}, 0505.
\url{http://doi.org/10.1088/1475-7516/2005/05/008}.

\bibitem{Easther:2006qu} 
Easther, R.; Kinney, W.H.; Powell, B.A.
The Lyth bound and the end of inflation.
\emph{JCAP} {\bf 2006}, \emph{0608}, 004.
\url{http://doi.org/doi:10.1088/1475-7516/2006/08/004}.


\bibitem{Garcia-Bellido:2014eva} 
Garcia-Bellido, J.; Roest, D.; Scalisi, M.; Zavala, I.
Can CMB data constrain the inflationary field range?
\emph{JCAP} {\bf 2014}, \emph{1409}, 006.
\url{http://doi.org/10.1088/1475-7516/2014/09/006}.

\bibitem{Garcia-Bellido:2014wfa} 
Garcia-Bellido, J.; Roest, D.; Scalisi, M.; Zavala, I.
Lyth bound of inflation with a tilt.
\emph{Phys.\ Rev.\ D } {\bf 2014}, \emph{90}, 123539.
\url{http://doi.org/0.1103/PhysRevD.90.123539}.

\bibitem{DiMarco:2017ihz} 
Di Marco, A.
Lyth Bound, eternal inflation and future cosmological missions.
\emph{Phys.\ Rev.\ D } {\bf 2017}, \emph{96}, 023511.
\url{http://doi.org/10.1103/PhysRevD.96.023511}.



\bibitem{Albrecht:1982mp}
Albrecht, A.; Steinhardt, P.J.; Turner, M.S.; Wilczek, F.
Reheating an Inflationary Universe.
\emph{Phys.\ Rev.\ Lett.\ } {\bf 1982}, \emph{48}, 1437.
\url{http://doi.org/10.1103/PhysRevLett.48.1437}.

\bibitem{Dolgov:1982th}
Dolgov, A.D.; Linde, A.D.
Baryon Asymmetry in Inflationary Universe.
\emph{Phys.\ Lett.\  }{\bf 1982}, \emph{116B}, 329.
\url{http://doi.org/10.1016/0370-2693(82)90292-1}.


\bibitem{Abbott:1982hn}
Abbott, L.F.; Farhi, E.; Wise, M.B.
Particle Production in the New Inflationary Cosmology.
\emph{Phys.\ Lett.\  }{\bf 1982}, \emph{117B}, 29.
\url{http://doi.org/10.1016/0370-2693(82)90867-X}.

\bibitem{Turner:1983he}
Turner, M.S.
Coherent Scalar Field Oscillations in an Expanding Universe.
\emph{Phys.\ Rev.\ D } {\bf 1983}, \emph{28}, 1243.
\url{http://doi.org/doi:10.1103/PhysRevD.28.1243}.

\bibitem{Shtanov:1993es}
Shtanov, Y.
Scalar-field dynamics and reheating of the universe in chaotic inflation scenario.
\emph{Ukr. Fiz. Zh.} \textbf{1993}, \emph{38}, 1425--1434.

\bibitem{DiMarco:2019czi}
Di Marco, A.; Gasperis, G.D.; Pradisi, G.; Cabella, P.
Energy Density, Temperature and Entropy Dynamics in Perturbative Reheating.
\emph{Phys. Rev. D} \textbf{2019}, \emph{100}, 123532.
\url{http://doi.org/10.1103/PhysRevD.100.123532}.

\bibitem{DiMarco:2021xzk}
Di Marco, A.; Pradisi, G.
Variable inflaton equation-of-state and reheating.
\emph{Int. J. Mod. Phys. A} \textbf{(2021}, \emph{36}, 2150095.
\url{http://doi.org/10.1142/S0217751X21500950}.


\bibitem{Dolgov:1989us} 
Dolgov, A.D.; Kirilova, D.P.
On Particle Creation by a Time Dependent Scalar Field.
\emph{Sov.\ J.\ Nucl.\ Phys.\ } {\bf 1990}, \emph{51}, 172.


\bibitem{Traschen:1990sw}
Traschen, J.H.; Brandenberger, R.H.
Particle Production During Out-of-equilibrium Phase Transitions.
\emph{Phys.\ Rev.\ D } {\bf 1990}, \emph{42}, 2491.
\url{http://doi.org/doi:10.1103/PhysRevD.42.2491}.


\bibitem{Kofman:1994rk}
Kofman, L.; Linde, A.D.; Starobinsky, A.A.
Reheating after inflation.
\emph{Phys.\ Rev.\ Lett.\ } {\bf 1994}, \emph{73}, 3195.
\url{http://doi.org/10.1103/PhysRevLett.73.3195}.


\bibitem{Shtanov:1994ce} 
Shtanov, Y.; Traschen, J.H.; Brandenberger, R.H.
Universe reheating after inflation.
\emph{Phys.\ Rev.\ D } {\bf 1995}, \emph{51}, 5438.
\url{http://doi.org/10.1103/PhysRevD.51.5438}.


\bibitem{Kofman:1997yn}
Kofman, L.; Linde, A.D.; Starobinsky, A.A.
Towards the theory of reheating after inflation.
\emph{Phys.\ Rev.\ D } {\bf 1997}, \emph{56}, 3258.
\url{http://doi.org/10.1103/PhysRevD.56.3258}.

\bibitem{Greene:1997fu}
Greene, P.B.; Kofman, L.; Linde, A.D.; Starobinsky, A.A.
Structure of resonance in preheating after inflation.
\emph{Phys. Rev. D} \textbf{1997}, \emph{56}, 6175--6192.
\url{http://doi.org/10.1103/PhysRevD.56.6175}.


\bibitem{Greene:1997ge}
Greene, B.R.; Prokopec, T.; Roos, T.G.
Inflaton decay and heavy particle production with negative coupling.
\emph{Phys.\ Rev.\ D } {\bf 1997}, \emph{56}, 6484.
\url{http://doi.org/10.1103/PhysRevD.56.6484}.

\bibitem{Greene:1998nh}
Greene, P.B.; Kofman, L.
Preheating of fermions.
\emph{Phys. Lett. B} \textbf{1999}, \emph{448}, 6--12.
\url{http://doi.org/10.1016/S0370-2693(99)00020-9}




\bibitem{Bassett:2005xm} 
Bassett, B.A.; Tsujikawa, S.; Wands, D.
Inflation dynamics and reheating.
\emph{Rev.\ Mod.\ Phys.\ } {\bf 2006}, \emph{78}, 537.
\url{http://doi.org/10.1103/RevModPhys.78.537}.


\bibitem{Frolov:2010sz} 
Frolov, A.V.
Non-linear Dynamics and Primordial Curvature Perturbations from Preheating.
\emph{Class.\ Quant.\ Grav.\ } {\bf 2010}, \emph{27}, 124006.
\url{http://doi.org/10.1088/0264-9381/27/12/124006}.

\bibitem{Allahverdi:2010xz}
Allahverdi, R.; Brandenberger, R.; Cyr-Racine, F.Y.; Mazumdar, A.
Reheating in Inflationary Cosmology: Theory and Applications.
\emph{Ann. Rev. Nucl. Part. Sci.  }\textbf{2010}, \emph{60}, 27--51.
\url{http://doi.org/10.1146/annurev.nucl.012809.104511}.

\bibitem{Amin:2014eta} 
Amin, M.A.; Hertzberg, M.P.; Kaiser, D.I.; Karouby, J.
Nonperturbative Dynamics of Reheating After Inflation: A Review.
\emph{Int.\ \mbox{J.\ Mod.}\ Phys.\ D } {\bf 2014}, \emph{24}, 1530003.
\url{http://doi.org/10.1142/S0218271815300037}.

\bibitem{Lozanov:2019jxc} 
Lozanov, K.D.
Lectures on Reheating after Inflation. \emph{arXiv} \textbf{2019},
arXiv:1907.04402.
\url{http://arxiv.org/abs/1907.04402}.



\bibitem{Belinsky:1985zd}
Belinsky, V.A.; Khalatnikov, I.M.; Grishchuk, L.P.; Zeldovich, Y.B.
Inflationary stages in cosmological models with a scalar field.
\emph{Phys. Lett. B} \textbf{1985}, \emph{155}, 232--236.
\url{http://doi.org/10.1016/0370-2693(85)90644-6}.

\bibitem{Guo:2003zf}
Guo, Z.K.; Piao, Y.S.; Cai, R.G.; Zhang, Y.Z.
Inflationary attractor from tachyonic matter.
\emph{Phys. Rev. D} \textbf{2003}, \emph{68}, 043508.
\url{http://doi.org/10.1103/PhysRevD.68.043508}.

\bibitem{Urena-Lopez:2007zal}
Urena-Lopez, L.A.; Reyes-Ibarra, M.J.
On the dynamics of a quadratic scalar field potential.
\emph{Int. J. Mod. Phys. D } \textbf{2009}, \emph{18}, 621--634.
\url{http://doi.org/10.1142/S0218271809014674}.


\bibitem{Remmen:2013eja}
Remmen, G.N.; Carroll, S.M.
Attractor Solutions in Scalar-Field Cosmology.
\emph{Phys. Rev. D} \textbf{2013}, \emph{88}, 083518.
\url{http://doi.org/10.1103/PhysRevD.88.083518}.





\bibitem{Ellis:1990wsa}
Ellis, G.F.R.; Madsen, M.S.
Exact scalar field cosmologies.
\emph{Class. Quant. Grav. }\textbf{1991}, \emph{8}, 667--676.
\url{http://doi.org/10.1088/0264-9381/8/4/012}.


\bibitem{Lidsey:1991zp}
Lidsey, J.E.
The Scalar field as dynamical variable in inflation.
\emph{Phys. Lett. B} \textbf{1991}, \emph{273}, 42--46.
\url{http://doi.org/10.1016/0370-2693(91)90550-A}.


\bibitem{Salopek:1990jq}
Salopek, D.S.; Bond, J.R.
Nonlinear evolution of long wavelength metric fluctuations in inflationary models.
\emph{Phys. Rev. D} \textbf{1990}, \emph{42}, 3936--3962.
\url{http://doi.org/10.1103/PhysRevD.42.3936}.

\bibitem{Muslimov:1990be}
Muslimov, A.G.
On the Scalar Field Dynamics in a Spatially Flat Friedman Universe.
\emph{Class. Quant. Grav. }\textbf{1990}, \emph{7}, 231--237.
\url{http://doi.org/10.1088/0264-9381/7/2/015}.


\bibitem{Barrow:1990vx}
Barrow, J.D.
Graduated inflationary universes.
\emph{Phys. Lett. B} \textbf{1990}, \emph{235}, 40--43.
\url{http://doi.org/10.1016/0370-2693(90)90093-L}.

\bibitem{Barrow:1993hn}
Barrow, J.D.
New types of inflationary universe.
\emph{Phys. Rev. D} \textbf{1993}, \emph{48}, 1585--1590.
\url{http://doi.org/10.1103/PhysRevD.48.1585}.


\bibitem{Barrow:1994nt}
Barrow, J.D.
Exact inflationary universes with potential minima.
\emph{Phys. Rev. D} \textbf{1994}, \emph{49}, 3055--3058.
\url{http://doi.org/10.1103/PhysRevD.49.3055}.

\bibitem{Barrow:1995xb}
Barrow, J.D.; Parsons, P.
Inflationary models with logarithmic potentials.
\emph{Phys. Rev. D} \textbf{1995}, \emph{52}, 5576--5587.
\url{http://doi.org/10.1103/PhysRevD.52.5576}.


\bibitem{Lidsey:1995np}
Lidsey, J.E.; Liddle, A.R.; Kolb, E.W.; Copeland, E.J.; Barreiro, T.; Abney, M.
Reconstructing the inflation potential : An overview.
\emph{Rev. Mod. Phys.} \textbf{1997}, \emph{69}, 373--410.
\url{http://doi.org/10.1103/RevModPhys.69.373}.


\bibitem{COBE:1992syq}
Smoot, G.F.; Bennett, C.L.; Kogut, A.; Wright, E.L.; Aymon, J.; Boggess, N.W.; Cheng, E.S.; De Amici, G.; Gulkis, S.; Hauser, M.G.; Hinshaw, G.
Structure in the COBE differential microwave radiometer first year maps.
\emph{Astrophys. J. Lett.} \textbf{1992}, \emph{396}, L1--L5.
\url{http://doi.org/10.1086/186504}.

\bibitem{Planck:2018jri}
Akrami, Y.; Arroja, F.; Ashdown, M.; Aumont, J.; Baccigalupi, C.; Ballardini, M.; B.; ay, A.J.; Barreiro, R.B.; Bartolo, N.; Basak, S.; Benabed, K.
Planck 2018 results. X. Constraints on inflation.
\emph{Astron. Astrophys. }\textbf{2020}, \emph{641}, A10.
\url{http://doi.org/10.1051/0004-6361/201833887}.

\bibitem{BICEP:2021xfz}
Ade, P.A.; Ahmed, Z.; Amiri, M.; Barkats, D.; Thakur, R.B.; Bischoff, C.A.; Beck, D.; Bock, J.J.; Boenish, H.; Bullock, E.; Buza, V.
Improved Constraints on Primordial Gravitational Waves using Planck, WMAP, and BICEP/Keck Observations through the 2018 Observing Season.
\emph{Phys. Rev. Lett. }\textbf{2021}, \emph{127}, 151301.
\url{http://doi.org/10.1103/PhysRevLett.127.151301}.





\bibitem{Liddle:2003as}
Liddle, A.R.; Leach, S.M.
How long before the end of inflation were observable perturbations produced?
\emph{Phys. Rev. D} \textbf{2003}, \emph{68}, 103503.
\url{http://doi.org/10.1103/PhysRevD.68.103503}.


\bibitem{German:2022sjd}
Germ\'an, G.; Quaglia, R.G.; Colorado, A.M.M.
Model independent bounds for the number of e-folds during the evolution of the universe.
\emph{JCAP} \textbf{2023}, \emph{03}, 004.
\url{http://doi.org/10.1088/1475-7516/2023/03/004}.




\bibitem{Das:2015uwa}
Das, K.; Dutta, K.; Maharana, A.
Inflationary Predictions and Moduli Masses.
\emph{Phys. Lett. B} \textbf{2015}, \emph{751}, 195--200.
\url{http://doi.org/10.1016/j.physletb.2015.10.041}.

\bibitem{Cicoli:2016olq}
Cicoli, M.; Dutta, K.; Maharana, A.; Quevedo, F.
Moduli Vacuum Misalignment and Precise Predictions in String Inflation.
\emph{JCAP} \textbf{2016}, \emph{08}, 006.
\url{http://doi.org/10.1088/1475-7516/2016/08/006}.

\bibitem{Maharana:2017fui}
Maharana, A.; Zavala, I.
Postinflationary scalar tensor cosmology and inflationary parameters.
\emph{Phys. Rev. D} \textbf{2018}, \emph{97}, 123518.
\url{http://doi.org/10.1103/PhysRevD.97.123518}.


\bibitem{DiMarco:2018bnw}
Di Marco, A.; Pradisi, G.; Cabella, P.
Inflationary scale, reheating scale, and pre-BBN cosmology with scalar fields.
\emph{Phys. Rev. D} \textbf{2018}, \emph{98}, 123511.
\url{http://doi.org/10.1103/PhysRevD.98.123511}.



\bibitem{Kawasaki:2000yn}
Kawasaki, M.; Yamaguchi, M.; Yanagida, T.
Natural chaotic inflation in supergravity.
\emph{Phys. Rev. Lett. }\textbf{2000}, \emph{85}, 3572--3575.
\url{http://doi.org/10.1103/PhysRevLett.85.3572}.


\bibitem{Takahashi:2010ky}
Takahashi, F.
Linear Inflation from Running Kinetic Term in Supergravity.
\emph{Phys. Lett. B} \textbf{2010}, \emph{693}, 140--143.
\url{http://doi.org/10.1016/j.physletb.2010.08.029}.


\bibitem{Nakayama:2010kt}
Nakayama, K.; Takahashi, F.
Running Kinetic Inflation.
\emph{JCAP} \textbf{2010}, \emph{11}, 009.
\url{http://doi.org/10.1088/1475-7516/2010/11/009}.

\bibitem{Nakayama:2010sk}
Nakayama, K.; Takahashi, F.
Higgs Chaotic Inflation in Standard Model and NMSSM.
\emph{JCAP} \textbf{2011}, \emph{02}, 010.
\url{http://doi.org/10.1088/1475-7516/2011/02/010}.

\bibitem{Nakayama:2010ga}
Nakayama, K.; Takahashi, F.
General Analysis of Inflation in the Jordan frame Supergravity.
\emph{JCAP} \textbf{2010}, \emph{11}, 039.
\url{http://doi.org/10.1088/1475-7516/2010/11/039}.

\bibitem{Harigaya:2012pg}
Harigaya, K.; Ibe, M.; Schmitz, K.; Yanagida, T.T.
Chaotic Inflation with a Fractional Power-Law Potential in Strongly Coupled Gauge Theories.
\emph{Phys. Lett. B} \textbf{2013}, \emph{720}, 125--129.
\url{http://doi.org/10.1016/j.physletb.2013.01.058}.


\bibitem{Silverstein:2008sg}
Silverstein, E.; Westphal, A.
Monodromy in the CMB: Gravity Waves and String Inflation.
\emph{Phys.\ Rev.\ D } {\bf 2008}, \emph{78}, 106003.
\url{http://doi.org/10.1103/PhysRevD.78.106003}.

\bibitem{McAllister:2008hb}
McAllister, L.; Silverstein, E.; Westphal, A.
Gravity Waves and Linear Inflation from Axion Monodromy.
\emph{Phys.\ Rev.\ D } {\bf 2010}, \emph{82}, 046003.
\url{http://doi.org/10.1103/PhysRevD.82.046003}.


\bibitem{McAllister:2014mpa} 
McAllister, L.; Silverstein, E.; Westphal, A.; Wrase, T.
The Powers of Monodromy.
\emph{JHEP} {\bf 2014}, \emph{1409}, 123.
\url{http://doi.org/10.1007/JHEP09(2014)123}.





\bibitem{Lucchin:1984yf}
Lucchin, F.; Matarrese, S.
Power Law Inflation.
\emph{Phys. Rev. D} \textbf{1985}, \emph{32}, 1316.
\url{http://doi.org/10.1103/PhysRevD.32.1316}.

\bibitem{Halliwell:1986ja}
Halliwell, J.J.
Scalar Fields in Cosmology with an Exponential Potential.
\emph{Phys. Lett. B} \textbf{1987}, \emph{185}, 341.
\url{http://doi.org/10.1016/0370-2693(87)91011-2}.

\bibitem{Burd:1988ss}
Burd, A.B.; Barrow, J.D.
Inflationary Models with Exponential Potentials.
\emph{Nucl. Phys. B} \textbf{1988}, \emph{308}, 929--945.
\url{http://doi.org/10.1016/0550-3213(88)90135-6}.





\bibitem{Kallosh:2013hoa}
Kallosh, R.; Linde, A.
Universality Class in Conformal Inflation.
\emph{JCAP} {\bf 2013}, \emph{1307}, 002.
\url{http://doi.org/10.1088/1475-7516/2013/07/002}.

\bibitem{Kallosh:2013daa}
Kallosh, R.; Linde, A.
Multi-field Conformal Cosmological Attractors.
\emph{JCAP} {\bf 2013}, \emph{1312}, 006.
\url{http://doi.org/10.1088/1475-7516/2013/12/006}.

\bibitem{Ferrara:2013rsa}
Ferrara, S.; Kallosh, R.; Linde, A.; Porrati, M.
Minimal Supergravity Models of Inflation.
\emph{Phys.\ Rev.\ D } {\bf 2013}, \emph{88},  085038.
\url{http://doi.org/10.1103/PhysRevD.88.085038}.


\bibitem{Kallosh:2013xya}
Kallosh, R.; Linde, A.
Superconformal generalizations of the Starobinsky model.
\emph{JCAP} {\bf 2013}, \emph{1306}, 028.
\url{http://doi.org/10.1088/1475-7516/2013/06/028}.

\bibitem{Kallosh:2013yoa}
Kallosh, R.; Linde, A.; Roest, D.
Superconformal Inflationary $\alpha$-Attractors.
\emph{JHEP} {\bf 2013}, \emph{1311}, 198.
\url{http://doi.org/10.1007/JHEP11(2013)198}.

\bibitem{Galante:2014ifa}
Galante, M.; Kallosh, R.; Linde, A.; Roest, D.
Unity of Cosmological Inflation Attractors.
\emph{Phys.\ Rev.\ Lett.\ } {\bf 2015}, \emph{114}, ,  141302.
\url{http://doi.org/10.1103/PhysRevLett.114.141302}.

\bibitem{Kallosh:2015zsa}
Kallosh, R.; Linde, A.
Escher in the Sky.
\emph{Comptes Rendus Phys.} \textbf{2015}, \emph{16}, 914--927.
\url{http://doi.org/10.1016/j.crhy.2015.07.004}.

\bibitem{Kallosh:2016gqp}
Kallosh, R.; Linde, A.
Cosmological Attractors and Asymptotic Freedom of the Inflaton Field.
\emph{JCAP} {\bf 2016}, \emph{1606},  047.
\url{http://doi.org/10.1088/1475-7516/2016/06/047}.

\bibitem{Maeda:1987xf}
Maeda, K.i.
Inflation as a Transient Attractor in R\textsuperscript{2} Cosmology.
\emph{Phys. Rev. D} \textbf{1988}, \emph{37}, 858.
\url{http://doi.org/10.1103/PhysRevD.37.858}.

\bibitem{Goncharov:1983mw}
Goncharov, A.B.; Linde, A.D.
Chaotic Inflation in Supergravity.
\emph{Phys. Lett. B} \textbf{1984}, \emph{139}, 27--30.
\url{http://doi.org/10.1016/0370-2693(84)90027-3}.

\bibitem{Goncharov:1984jlb}
Goncharov, A.S.; Linde, A.D.
Chaotic inflation of the universe in supergravity.
\emph{Sov. Phys. JETP} \textbf{1984}, \emph{59}, 930--933.







\bibitem{Pradisi:2022nmh}
Pradisi, G.; Salvio, A.
(In)equivalence of metric-affine and metric effective field theories.
\emph{Eur. Phys. J. C} \textbf{2022}, \emph{82}, 840.
\url{http://doi.org/10.1140/epjc/s10052-022-10825-9}.

\bibitem{Salvio:2022suk}
Salvio, A.
Inflating and reheating the Universe with an independent affine connection.
\emph{Phys. Rev. D} \textbf{2022}, \emph{106}, 103510.
\url{http://doi.org/10.1103/PhysRevD.106.103510}.

\bibitem{DiMarco:2023ncs}
Di Marco, A.; Orazi, E.; Pradisi, G.
Einstein\textendash{}Cartan pseudoscalaron inflation.
\emph{Eur. Phys. J. C }\textbf{2024}, \emph{84}, 146.
\url{http://doi.org/10.1140/epjc/s10052-024-12482-6}.



\bibitem{Cicoli:2008va}
Cicoli, M.; Conlon, J.P.; Quevedo, F.
General Analysis of LARGE Volume Scenarios with String Loop Moduli Stabilisation.
\emph{JHEP} \textbf{2008}, \emph{10}, 105.
\url{http://doi.org/10.1088/1126-6708/2008/10/105}.

\bibitem{Cicoli:2008gp}
Cicoli, M.; Burgess, C.P.; Quevedo, F.
Fibre Inflation: Observable Gravity Waves from IIB String Compactifications.
\emph{JCAP} \textbf{2009}, \emph{03}, 013.
\url{http://doi.org/10.1088/1475-7516/2009/03/013}.

\bibitem{Cicoli:2016xae}
Cicoli, M.; Muia, F.; Shukla, P.
Global Embedding of Fibre Inflation Models.
\emph{JHEP} \textbf{2016}, \emph{11}, 182.
\url{http://doi.org/10.1007/JHEP11(2016)182}.

\bibitem{Cicoli:2020bao}
Cicoli, M.; Valentino, E.D.
Fitting string inflation to real cosmological data: The fiber inflation case.
\emph{Phys. Rev. D} \textbf{2020}, \emph{102}, 043521.
\url{http://doi.org/10.1103/PhysRevD.102.043521}.



\bibitem{Acquaviva:2002ud}
Acquaviva, V.; Bartolo, N.; Matarrese, S.; Riotto, A.
Second order cosmological perturbations from inflation.
\emph{Nucl. Phys. B} \textbf{2003}, \emph{667}, 119--148.
\url{http://doi.org/10.1016/S0550-3213(03)00550-9}.

\bibitem{Maldacena:2002vr}
Maldacena, J.M.
Non-Gaussian features of primordial fluctuations in single field inflationary models.
\emph{JHEP} \textbf{2003}, \emph{05}, 013.
\url{http://doi.org/10.1088/1126-6708/2003/05/013}.

\bibitem{Bartolo:2004if}
Bartolo, N.; Komatsu, E.; Matarrese, S.; Riotto, A.
Non-Gaussianity from inflation: Theory and observations.
\emph{Phys. Rept.} \textbf{2004}, \emph{402}, 103--266.
\url{http://doi.org/10.1016/j.physrep.2004.08.022}.

\bibitem{Babich:2004gb}
Babich, D.; Creminelli, P.; Zaldarriaga, M.
The Shape of non-Gaussianities.
\emph{JCAP} \textbf{2004}, \emph{08}, 009.
\url{http://doi.org/10.1088/1475-7516/2004/08/009}.

\bibitem{Chen:2010xka}
Chen, X.
Primordial Non-Gaussianities from Inflation Models.
\emph{Adv. Astron.} \textbf{2010}, \emph{2010}, 638979.
\url{http://doi.org/10.1155/2010/638979}.



\bibitem{Arkani-Hamed:2015bza}
Arkani-Hamed, N.; Maldacena, J.
Cosmological Collider Physics. \emph{arXiv} \textbf{2015},
arXiv:1503.08043.
\url{http://arxiv.org/abs/1503.08043}.

\bibitem{Sasaki:1995aw}
Sasaki, M.; Stewart, E.D.
A General analytic formula for the spectral index of the density perturbations produced during inflation.
\emph{Prog. Theor. Phys. }\textbf{1996}, \emph{95}, 71--78.
\url{http://doi.org/10.1143/PTP.95.71}.


\bibitem{Lyth:2004gb}
Lyth, D.H.; Malik, K.A.; Sasaki, M.
A General proof of the conservation of the curvature perturbation.
\emph{JCAP} \textbf{2005}, \emph{05}, 004.
\url{http://doi.org/10.1088/1475-7516/2005/05/004}.

\bibitem{Wands:2000dp}
Wands, D.; Malik, K.A.; Lyth, D.H.; Liddle, A.R.
A New approach to the evolution of cosmological perturbations on large scales.
\emph{Phys. Rev. D} \textbf{2000}, \emph{62}, 043527.
\url{http://doi.org/10.1103/PhysRevD.62.043527}.

\bibitem{Dias:2012qy}
Dias, M.; Ribeiro, R.H.; Seery, D.
The \ensuremath{\delta}N formula is the dynamical renormalization group.
\emph{JCAP} \textbf{2013}, \emph{10}, 062.
\url{http://doi.org/10.1088/1475-7516/2013/10/062}.


\bibitem{Sugiyama:2012tj}
Sugiyama, N.S.; Komatsu, E.; Futamase, T.
$\delta$N formalism.
\emph{Phys. Rev. D} \textbf{2013}, \emph{87}, 023530.
\url{http://doi.org/10.1103/PhysRevD.87.023530}.

\bibitem{Abolhasani:2019cqw}
Abolhasani, A.A.; Firouzjahi, H.; Naruko, A.; Sasaki, M.
\emph{Delta N Formalism in Cosmological Perturbation Theory};
WSP: Phoenix, AZ, USA, 2019.
\url{http://doi.org/10.1142/10953}.




\end{thebibliography}
\newcommand{\journal}[2]{\href{http://dx.doi.org/#1}{#2}}
\newcommand{\arxiv}[2]{\href{http://arxiv.org/abs/#1}{[arxiv:#1 #2]}}

\end{document}